\documentclass[preprint,showpacs]{revtex4}
\usepackage{graphicx}
\usepackage{dcolumn}
\usepackage{amsmath}
\usepackage{epstopdf}
\usepackage[colorlinks=true, citecolor=red, urlcolor = red, linkcolor= blue, bookmarks=true]{hyperref}

\begin{document}
\title{$5D$ rotating regular Myers-Perry black holes and their shadow}
\author{Fazlay Ahmed$^{1,\;2}$} \email{fazleyamuphysics@gmail.com}
\author{Dharm Veer Singh $^{1,\;3}$} \email{veerdsingh@gmail.com}
\author{Sushant G. Ghosh$^{1, \;4}$}\email{sghosh2@jmi.ac.in}
\affiliation{$^1$Center for Theoretical Physics, Jamia Millia Islamia, New Delhi 110025, India}
\affiliation{$^2$Department of Physics, PDM University, Bahadurgarh 124507, Haryana, India}
\affiliation{$^3$Department of Physics, Institute of Applied Sciences and Humanities, GLA University, Mathura 281406, Uttar Pradesh, India}
\affiliation{$^4$Astrophysics and Cosmology Research Unit, School of Mathematics, Statistics and Computer Science, University of KwaZulu-Natal, Private Bag X54001,
 Durban 4000, South Africa}

 \begin{abstract}
It is widely believed that the curvature singularities are an artifact of general relativity and may not exist in the Universe, and likely to be defined by the quantum gravity. In the absence of a successful quantum gravity, significant attention has been shifted towards the regular models. We present a five-dimensional ($5D$) rotating regular Myers-Perry like black hole metric with two rotation parameters ($a,b$) and a free parameter ($k$). Our $5D$ rotating regular metric encompasses $5D$ Myers-Perry black hole which can be recovered as a special case when $k=0$. The size of the shadow cast by the black holes is significantly smaller and more distorted when the parameter $k$ is increased.    
\end{abstract}

\maketitle

\section{Introduction}
Black holes, solutions of pure gravity equations $R_{ab}=0$, are one of the most extraordinary predictions in four-dimensional General Relativity and have been the subject of intense research since their inception (see e.g. the reviews \cite{Horowitz:2012nnc,Emparan:2008eg}). One of the most important results on four-dimensional black holes is the uniqueness theorem \cite{Hawking:1971vc}, stating that an asymptotically flat black hole in vacuum $4D$ general relativity with a regular horizon is described by the Kerr metric \cite{Kerr:1963ud}. The Kerr metric when perturbed ultimately leads to another Kerr metric with slightly different mass and angular momentum thereby stable under a small perturbation \cite{Whiting:1988vc}. However, in recent years, motivated by ideas in Brane-world cosmology and string theory, attention has been shifted to the black holes in higher dimensions as a growing volume of recent literature indicates \cite{Emparan:2001wn,Myers:1986un,Horowitz:2011cq,Randall:1999ee,Reall:2002bh}. The physics of higher-dimensional black holes can be strikingly different, and much richer, than in four dimensions \cite{Emparan:2008eg}. Also, the statistical counting of black hole entropy was first accomplished for a five-dimensional ($5D$) black hole in string theory \cite{Strominger:1996sh}. The $D$-dimensional black holes are related to those of a quantum field theory in $(D-1)$ dimensions \cite{Aharony:1999ti} through AdS/CFT correspondence.\\ 
The first higher dimensional black hole solution was obtained by Schwarzschild-Tangherlini \cite{Tangherlini:1963bw} whereas the Myers-Perry black holes \cite{Myers:1986un} are the higher-dimensional generalizations of the Kerr black hole \cite{Kerr:1963ud}. In particular, the $5D$ Myers-Perry black hole corresponds to a rotating black hole which is an asymptotically flat stationary solution of the $R_{ab}=0$ with an event horizon that has the spherical topology $S_3$. 
The metric of the $5D$ Myers-Perry black hole, in Boyer-Lindquist coordinates $(t,r,\theta,\phi,\psi)$ reads
\begin{eqnarray}\label{5mpbh}
ds^2 &=& -\left(1-\frac{M}{\rho^2} \right)dt^2 + \frac{r^2 \rho^2}{\Delta} +\rho^2 d \theta^2 -\frac{2 a M \sin^2\theta}{\rho^2} dt d\phi  - \frac{2 b M \cos^2\theta}{\rho^2} dt d\psi \nonumber\\ && + \frac{2 a b M \sin^2 \theta \cos^2 \theta}{\rho^2} d\phi d\psi  + \sin^2\theta \left(r^2+a^2+\frac{M a^2 \sin^2 \theta}{\rho^2} \right)d\phi^2 \nonumber\\ && + \cos^2\theta \left(r^2+b^2+\frac{M b^2 \cos^2\theta}{\rho^2} \right)d\psi^2,
\end{eqnarray}
where
\begin{eqnarray}\label{5delta}
\Delta = (r^2+a^2)(r^2+b^2)-M r^2,  \quad \quad   \rho^2 = r^2+a^2 \cos^2\theta + b^2 \sin^2\theta.
\end{eqnarray}
The metric~(\ref{5mpbh}) is characterized by its mass $(M)$ and two angular momenta ($ a,\; b $), and it exhibit instability when one of the two angular momenta is large \cite{Harmark:2004rm}. The two angles obey $0 \leq \phi \leq 2 \pi$ and $0\leq \psi \leq 2\pi$. The metric~(\ref{5mpbh}) is stationary with timelike Killing field $\partial/{\partial t}$, and it is an axis-symmetric to two rotational axes, corresponding to spacelike Killing field $\partial/{\partial \phi}$ and $\partial/{\partial \psi}$.
The metric~(\ref{5mpbh}), like Kerr black holes, have horizons given by the positive root of
\begin{eqnarray}
\Delta=(r^2+a^2)(r^2+b^2)-M r^2,
\end{eqnarray}
which is explicitly given by \cite{Frolov:2003en}
\begin{eqnarray}
r_{\pm}^{2} &=& \frac{1}{2}\Big(M-(a^2+b^2) \pm \sqrt{(M-(a^2+b^2))^2-4 a^2b^2}\Big),
\end{eqnarray}
where $r_+$ and $r_-$, respectively, denote the outer and inner horizons. The solution~(\ref{5mpbh}) describes non-extremal black holes for $r_+ > r_-$, and when $r_+ = r_-$, it represents an extremal black hole. If $(M-(a^2+b^2))^2<4a^2b^2$, no horizon exist, which defines a region in $(a,b)$ where (\ref{5mpbh}) represents a naked singularity. When $a=b$, the $5D$ Myers-Perry black hole is extremal at $M=4a^2$, while naked singularity occurs when $M<4a^2$, and $M>4a^2$ correspond to black hole with two horizons.
The Killing horizon corresponds to
\begin{eqnarray}
\frac{\partial}{\partial t}+\Omega_{\phi}\frac{\partial}{\partial \phi}+ \Omega_{\psi}\frac{\partial}{\partial \psi},
\end{eqnarray}
where $\Omega_{\phi}$ and $\Omega_{\psi}$ represent angular velocities with respect to the axes $\frac{\partial}{\partial \phi}$ and $\frac{\partial}{\partial \psi}$, given by
\begin{eqnarray}
\Omega_{\phi} = \frac{a}{r_{+}^2 + a^2} \quad \mbox{and} \quad \Omega_{\psi} = \frac{b}{r_{+}^2 + b^2}.
\end{eqnarray}
It is important to note, like Kerr black hole, the metric (\ref{5mpbh}) has a curvature singularity when $\rho^2=0$ regions where the theory fails to be predictive. The exact nature of such a singularity might be resolved by a successful theory of quantum gravity, which may become predictive in the extreme region where curvature diverges. In the absence of well-defined quantum gravity, the attention has been shifted to the nonsingular or regular models of black holes where the central singularity is replaced with a de Sitter core or regular spacetime region to make dynamics modified, thereby the classical theory to work and the spacetime curvature is well defined in this regular region \cite{Frolov:1981mz,Frolov:1998wf}. It was Bardeen \cite{Bardeen:1968} who realized the idea of a central matter replacing the singularity by a regular de Sitter core to propose the first regular black hole, and the physical source associated to Bardeen solution was obtained by Ay\'{o}n-Beato and Garc\'{i}a \cite{AyonBeato:1998ub}. They invoked nonlinear electrodynamics to generate the Bardeen model as an exact nonlinear magnetic monopole, also suggested regular black holes
from nonlinear electric fields \cite{ABG1}, which goes exactly encompasses the Reisnner-Nordstrom black holes as special case. Bronnikov \cite{Bronnikov:2000vy} proposed several regular black holes in which source are the fields, the core is an expanding universe with de Sitter asymptotes and the exterior outer region tends to Schwarzschild black hole. Subsequently, This idea triggered intense activities in the investigation of many other regular black holes \cite{Dymnikova:1992ux,AyonBeato:1999rg,Lemos:2011dq,Hayward:2005gi,Balart:2014jia, dvs2019,ads}, including the rotating ones \cite{Bambi:2013ufa,Toshmatov:2014nya,Ghosh:2014pba} generalizing the Kerr black holes. \\
The main purpose of this paper is to obtain a $5D$ regular or nonsingular rotating black hole, a generalization of the $5D$ Myers-Perry black holes metric~(\ref{5mpbh}) without curvature singularity and examine its properties. We have also studied the shadow of the black hole, which is an image of the photon sphere (at $r=3M$ for Schwarzschild black hole) \cite{Perlick:2015vta,Cunha:2018gql}. The first step towards the study of a black hole shadow was done by Bardeen \cite{bardeen2}, who calculated its for the Kerr black hole. The shadow of a Kerr black hole is not a perfect circle but for Schwarzschild black hole, its has a perfect circle shape \cite{Synge:1966okc}. Later the study of the black hole shadow has been extended for many other black holes \cite{Bambi:2008jg,Hertog:2019hfb,Amarilla:2011fxx,Cunha:2015yba}, including the regular ones \cite{Abdujabbarov:2016hnw,Amir:2016cen}.\\
The paper is organized as follows. The Sec.~\ref{5Drbh} is devoted to 
construct a $5D$ rotating regular black holes with two angular momenta, using the modified Newman-Janis algorithm. The properties of a $5D$ rotating regular solution are a subject of the Sec.~\ref{5Dprp}. How the horizon structure of the black hole is effected by charge is the subject of the Sec.~\ref{rrbh}, and also a detailed analysis of ergoregion is carried out. The equations of motion in the background of $5D$ rotating regular Myers-Perry black hole are derived in the Sec.~\ref{pmbh}, and exact expressions for the impact parameters, which determine the black hole shadow, are also obtained. The Sec.~\ref{pmbh} is devoted to detailed investigation of the shadow cast by the $5D$ rotating regular Myers-Perry black holes alongwith a discussion on energy emission rate. We end the paper with concluding remarks in the Sec.~\ref{cnbh}. We use the natural units, i.e., $G=c=1$.

\section{5D rotating regular Myers-Perry black holes}\label{5Drbh}
The nonlinear electrodynamics theory is governed by the action
\begin{equation}\label{action1}
S= \int d^5 x \sqrt{-g} \left(R - \mathcal{L}(F)\right),
\end{equation}
where $R$ is the Ricci scalar, $g$ is the determinant of the metric $g_{ab}$ and $\mathcal{L}(F)$ is the lagrangian density with $F = F_{ab} F^{ab}/4$, where $F_{ab}$ is the electromagnetic field strength. \\
Varying the action (\ref{action1}), result into the following equations of motion
\begin{eqnarray}\label{Rab}
R_{ab} - \frac{1}{2} g_{ab} = T_{ab} \equiv \left[\frac{\partial \mathcal{L} (F)}{\partial F} F_{ac} F^{c}_{b}-g_{ab} \mathcal{L}(F) \right],\nonumber\\
 \nabla_{a}\left(\frac{\partial {\cal{L}}(F)}{\partial F}F^{a b}\right)=0\qquad \text{and} \qquad \nabla_{\mu}(* F^{ab})=0,
\end{eqnarray}
with $R_{ab}$ is the Ricci tensor. The Lagrangian density $\mathcal{L}(F)$ is given by \cite{Ghosh:2018bxg}
\begin{equation}\label{lf}
\mathcal{L}(F)=3Fe^{-\frac{k}{e}(2eF)^{\frac{1}{3}}}.
\end{equation}
For spherically symmetric spacetimes, the  nonvanishing components of $F_{ab}$ are  $F_{\theta\phi}, F_{\theta\psi}$ and $F_{\phi\psi}$. The Maxwell field for nonlinear electrodynamics reads
\begin{equation}
F_{ab}=2\delta^{\theta}_{[a}\delta^{\phi}_{b]}Z(r,\theta,\phi),
\label{8}
\end{equation}
where $Z$ has been suitably modified for the $5D$. Substituted Eq. (\ref{8}) into  Eq. (\ref{Rab}) and integrating, we obtain 
\begin{equation}
F_{ab}=2\delta^{\theta}_{[a}\delta^{\phi}_{b]}e(r) \sin^2\theta\sin\phi.
\label{9}
\end{equation}
Eq. (\ref{9}) implies $dF=0$, thereby $ e^{\prime}(r)\sin^2\theta\sin\phi dr\wedge d\theta\wedge d\phi\wedge d\psi=0$.
This leads to $e(r)= e$ = constant. Interestingly, the other components of $F_{ab}$ have negligible influence in comparison to $F_{\theta\phi}$ \cite{Hendi:2017phi, Panahiyan:2018gzq}.  Hence the field strength can be simplified to
\begin{equation}
F_{\theta\phi}=\frac{e}{r}\sin\theta,\qquad \text{and} \qquad  F=\frac{e^2}{2r^6}.
\label{10}
\end{equation}
Using Eq. (\ref{10})  in Eq. (\ref{lf}), we obtain
\begin{equation}
 {\cal{L(F)}}=\frac{3 e^2 }{r^6}e^{-e^2/{M r^2}}. 
\label{11}
\end{equation}
where the free parameter $k$ is related to charge $e$ via $e^2=Mk$.
The components of energy-momentum tensor are written as
\begin{eqnarray}\label{energymoment}
&& T^t_t=T^r_r= \rho = \frac{3 M k }{r^6}e^{-k/r^2}, \nonumber\\ 
&& T^{\theta}_{\theta}=T^{\phi}_{\phi}=T^{\psi}_{\psi}=\frac{Mk}{r^6}\left(\frac{2k}{r^2}-3\right)e^{-k/r^2}.
\end{eqnarray}
The general metric for $5D$ static and spherically symmetric spacetime has the following form
{\begin{eqnarray}\label{genmetric}
ds^2 = -f(r)dt^2 + \frac{1}{h(r)}dr^2+r^2 d\Omega_3^2,
\end{eqnarray}
where $f(r)$ and $h(r)$ are the metric functions and $d\Omega_3^2=d\theta^2+\sin^2\theta d\phi^2+\cos^2 \theta d\psi^2$ is the metric of $3D$ sphere. Now we assume $h(r)=f(r)$.
It turns out that the Eqs.~(\ref{Rab}), (\ref{lf}), and (\ref{energymoment}) alongwith Eq.~(\ref{genmetric}) admit the solution \cite{Ghosh:2018bxg} 
\begin{eqnarray}
f(r)=1-\frac{ M }{r^2}\,e^{-k/ r^2}.
\end{eqnarray}
Now, the metric for the $5D$ spherically symmetric regular black hole takes the form
\begin{equation}\label{nonrot}
ds^2 = -\Big(1-\frac{ M e^{-k/ r^2}}{r^2}\Big)dt^2 + \frac{1}{\Big(1-\frac{ M e^{-k/ r^2}}{r^2}\Big)} dr^2+r^2 d\Omega_3^2,
\end{equation}
 Clearly, the metric~(\ref{nonrot}) reduces to $5D$ Schwarzschild-Tengerhalini solution \cite{Tangherlini:1963bw} as a special case when $k=0$, and the $5D$ Minkowski solution for $M=0$. 
Further, the metric (\ref{nonrot}) when $r>>k$ becomes
\begin{equation}
d s^2 = \Big(1-\frac{M}{r^2}+\frac{e^2}{r^4}\Big)dt^2 + \Big(1-\frac{M}{r^2}+\frac{e^2}{r^4}\Big)^{-1} dr^2+r^2 d\Omega^2_3,
\end{equation}
which is $5D$ Reissner-N$\ddot{o}$rdstorm solution. Thus, we have a new exact $5D$ black hole with a free parameter $k$, that measures deviation from the $5D$  Schwarzschild-Tengerhalini black holes. It is seen that solution (\ref{nonrot}), for a range of values of parameters, is regular everywhere including at $r=0$.
\begin{figure}
   \includegraphics[scale=0.75]{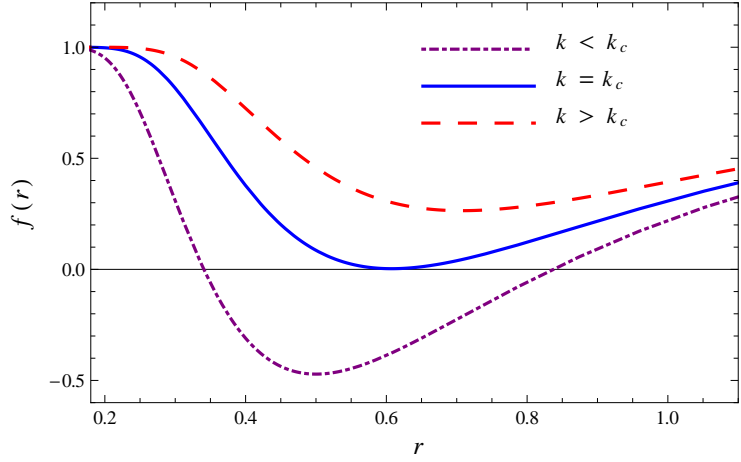}
         \caption{\label{frr} Plots showing the behavior of metric function $f(r)$ vs radius $r$.}
\end{figure}
We write 
\begin{equation}
f^{\prime}(r)=\frac{2M}{r^3}\left(1-\frac{k}{r^2}\right)e^{-k/r^2},
\end{equation}
which has roots at $r^2=k$, and the metric function gives it minimum value at $r^2=k$, which is
\begin{equation}
f_{min}=f(k)=1-\frac{M}{k e}.
\end{equation}
It turns out that $f(k)<0$, $f(k_c)=0$, and $f(k)>0$, respectively, when $k<M/e$, $k=k_c=M/e$, and $k>M/e$. The three cases corresponding, respectively to black hole with two horizons, extremal black holes with degenerate horizons and no black hole (cf. Fig.~\ref{frr}). \\
 The 5-acceleration, $a^b=u^a \nabla_a u^b$ is given by
\begin{equation}
a^b=\left(0, \frac{M (1-\frac{k}{r^2})}{r^3}e^{-k/r^2},0,0,0 \right),
\end{equation}
where $u^b$ is a $5$-velocity.
$a^b$ vanishes at $r^2=k$, where the function $f(r)$ is minimum. The gravitational field becomes repulsive for $r^2<k$, where acceleration is negative. 
The surface gravity $\kappa$ at the horizon is given by 
\begin{equation}
\kappa = \sqrt{a^b a_b} \sqrt{-g_{tt}}|_{+}=\frac{1}{r_+}\left(1-\frac{k}{r_+^2}\right).
\end{equation}
$\kappa$ or temperature vanishes at $r^2=k$, where the black hole is extremal.\\ 
Now, we would like to drive a rotating counterpart of solution~(\ref{nonrot}).
The Newman Janis algorithm (NJA) is a step by step procedure of complex transformation to generate new rotating black hole solution from the spherically symmetric counterpart, as Kerr/Kerr-Newman can be exactly generated from the Schwarzschild/Reissner-N$\ddot{o}$rdstorm metrics \cite{Newman:1965tw}. The NJA provides exact solutions \cite{Kim:1998iw,Drake:1997hh}. However, they do not always provide an exact solution in modified gravity, but may give useful insights into the structure of the solutions.  Erbin and Heurtier \cite{Erbin:2014lwa} propose an extension of NJA to $5D$ with two angular momenta, using the recipe of Giampieri \cite{Giampieri} to generate $5D$ Myers-Perry black holes \cite{Myers:1986un}. Hence, we apply the procedure in Refs. \cite{Giampieri,Erbin:2014lwa} to generate $5D$ rotating regular black holes starting with the spherically symmetric black holes (\ref{nonrot}), with only requirement to recover the Myers-Perry black hole metric for $k=0$, and exponential term is left as it is. \\
Let us start with the $5D$ spherically symmetric regular black hole metric rewritten as
\begin{eqnarray}\label{frmetric}
d s^2 = -f(r) dt^2 + f(r)^{-1} dr^2+r^2 d\Omega^2_3.
\end{eqnarray}
In the $5D$, the planes that can be made rotation, are $(r,\phi)$ and $(r,\psi)$. 
The first step is to rewrite the metric~(\ref{frmetric}) in null coordinates as
\begin{eqnarray}\label{metric1}
ds^2 &=& -du(du+2dr)+(1-f(r))du^2+r^2\left(d\theta^2+\sin^2\theta d\phi^2\right) + R^2\cos^2\theta d\psi^2,
\end{eqnarray}
where $R(r)=Re(r)$. Let us go for the first complex transformation, related to $(r,\phi)$-plane as \cite{Erbin:2014lwa}
\begin{eqnarray}\label{complex1}
&& u=u^{\prime} +ia \cos\chi_1,\;\;\; \quad \; r=r^{\prime} -ia \cos\chi_1,\nonumber\\
&& id\chi_1= \sin\chi_1 d\phi,\;\;\;\quad \quad \; \mbox{with}\;\;\chi_1=\theta,\nonumber\\
&& du=du^{\prime} -a \sin^2\theta d\phi,\;\;\; dr=dr^{\prime} +a \sin^2\theta d\phi,
\end{eqnarray}
with $f(r)$ to be replaced by ${f}^{(1)}={f}^{(1)}(r,\theta)$, as $f(r)$ must be complexified twice as compared to once in the original NJA \cite{Newman:1965tw}. The transformation (\ref{complex1}) on (\ref{metric1}) result into
\begin{eqnarray}\label{metric2}
ds^2&=&-du^2-2dudr+\left(1-{f}^{(1)}\right)\left(du-a\sin^2\theta d\phi\right)^2  + 2a\sin^2\theta dr d\phi  +\left(r^2+a^2\cos^2\theta\right)d\theta^2 \nonumber\\&& + \left(r^2+a^2\right)\sin^2\theta d\phi^2 +r^2\cos^2\theta d\psi^2.
\end{eqnarray}
where $R(r)=r^{\prime}$ and we have omitted the primes,
and the function ${f}^{(1)}$ reads
\begin{eqnarray}\label{metricf1}
{f}^{(1)}=1-\frac{  M e^{-k/r^2}}{|r|^2}=1-\frac{ M e^{-k/r^2}}{r^2+a^2\cos^2\theta}
\end{eqnarray}
Now, making a complex transformation in $(r,\psi)$-plane, as follows
\begin{eqnarray}\label{complex2}
&& u=u^{\prime} +ib\cos\chi_2,\;\;\; \quad \;\; r=r^{\prime}-ib\cos\chi_2,\nonumber\\
&& id\chi_2=-\cos\chi_2 d\psi,\;\;\; \quad  \mbox{with}\;\;\chi_2=\theta,\nonumber\\
&& du=du^{\prime}-b\cos^2\theta d\psi, \;\;\; dr=dr^{\prime}+b\cos^2\theta d\psi,
\end{eqnarray}
with this transformation, the metric takes the form
\begin{eqnarray}\label{metric3}
ds^2&=&-du^2-2dudr+\left(1-{f}^{(2)}\right)\left(du-a\sin^2\theta d\psi\right)^2  +2a\sin^2\theta dR d\phi+\rho^2 d\theta^2+\nonumber\\&&\left(R^2+a^2\right)\sin^2\theta d\phi^2+r^2\cos^2\theta d\psi^2,
\end{eqnarray}
with once again using the function $R(r)=Re(r)$. Finally, the metric~(\ref{metric3}) can be written as
 \begin{eqnarray}\label{metric4}
ds^2&=&-du^2-2dudr + \rho^2 d\theta^2 + \left(1-{f}^{ (1,2) }\right)\left(du-a\sin^2\theta d \phi - b \cos^2\theta d\psi\right)^2   +2a \sin^2\theta dr d \phi \nonumber\\ && + \left(r^2+a^2\right)\sin^2\theta d\phi^2   + 2 b \cos^2\theta dr d\psi +\left(r^2+b^2\right)\cos^2\theta d\psi^2,
\end{eqnarray}
with once again, $R=r^{\prime}$ and omitting primes, 
where 
\begin{eqnarray}\label{rho}
\rho^2=r^2+a^2\cos^2\theta+b^2\sin^2\theta,
\end{eqnarray}
and the function ${f}^{ (1,2)}$ after two successive complexification takes the form
\begin{eqnarray}\label{metrcf2}
 {f}^{(1,2)} = 1-\frac{  M e^{-k/r^2}}{|r|^2+a^2\cos^2\theta} = 1-\frac{M e^{-k/r^2}}{r^2+a^2\cos^2\theta+b^2\sin^2\theta} = 1-\frac{  M e^{-k/r^2}}{\rho^2}.
\end{eqnarray}
The metric (\ref{metric4}) can be transformed to Boyer-Lindquist coordinates to obtain \cite{Erbin:2014lwa}
 \begin{eqnarray}\label{metric5}
ds^2 &= &- \Big(1-\frac{M e^{-k/r^2}}{\rho^2}\Big) dt^2-\frac{2 M a \sin^2 \theta e^{-k/r^2}}{\rho^2} dt d\phi  - \frac{2 M b \cos^2 \theta e^{-k/r^2}}{\rho^2} dt d\psi \nonumber\\&& + \frac{r^2 \rho^2}{\Delta} dr^2 + \rho^2 d\theta^2  + \sin^2 \theta \Big(r^2 + a^2 +\frac{M a^2 \sin^2 \theta e^{-k/r^2}}{\rho^2}\Big) d\phi^2 \nonumber\\ && + \frac{2 M a b \sin^2 \theta \cos^2 \theta e^{-k/r^2}}{\rho^2} d\phi d\psi + \cos^2 \theta \Big(r^2 + b^2 +\frac{M b^2 \cos^2 \theta e^{-k/r^2}}{\rho^2}\Big) d \psi^2,
\end{eqnarray}
where $\Delta$ is defined as
\begin{eqnarray}
\Delta=\left (r^2+a^2\right)\left(r^2+b^2\right)-  M r^2 e^{-k/r^2}.
\end{eqnarray}
This metric~(\ref{metric5}), describes the $5D$ rotating regular Myers-Perry black hole with two angular momenta $a$ and $b$. One recovers the standard $5D$ rotating Myers-Perry black hole \cite{Myers:1986un}, when $k=0$ and spherically symmetric (\ref{nonrot}), when $a=b=0$. In addition, if $k=0$, then metric~(\ref{metric5}) becomes Schwarzschild-Tengerhlani solution \cite{Tangherlini:1963bw}. When $M=0$, the metric~(\ref{metric5}) becomes flat in oblate bipolar coordinates \cite{Frolov:2003en}. We have generated the metric~(\ref{metric5}) via the Newman-Janis algorithm, which is widely used to generate rotating black hole solutions (see refs. \cite{Newman:1965tw,Kim:1998iw,Drake:1997hh,Erbin:2014lwa,Giampieri}) including the regular black hole metrics (see \cite{Bambi:2013ufa,Ghosh:2014pba}). It turns out that the Newman-Janis algorithm works fine when the spherically symmetric metric is vacuum and in this case, the rotating counterpart is also vacuum. But for the present case, the rotating metric~(\ref{metric5}) may have additional stresses, but they fall quite rapidly. It may be also mention that the metric~(\ref{metric5}) has all correct limits.

\section{Properties of the Solution}\label{5Dprp}

The metric~(\ref{metric5}) is invariant under the simultaneous inversion of the time coordinate $t \rightarrow -t$ and the angles $\phi \rightarrow -\phi$, $\psi \rightarrow -\psi$, and under the transformation of $a \leftrightarrow b$, $\phi  \leftrightarrow \psi$ and $\theta  \leftrightarrow \frac{\pi}{2}-\theta$.
  It implies the existence of three Killing vectors
 \begin{eqnarray}\label{killing}
 \xi_{(t)}=\frac{\partial}{\partial t}, \quad \quad \xi_{(\phi)}=\frac{\partial}{\partial \phi}, \quad \quad \xi_{(\psi)}=\frac{\partial}{\partial \psi}.
 \end{eqnarray}
and we have
\begin{eqnarray}
\xi_{t}\cdot \xi_{t} &=& g_{tt} = -1 + \frac{M e^{-k/r^2}}{\rho^2}, \nonumber\\
\xi_{\phi}\cdot \xi_{\phi} &=& g_{\phi \phi} = \sin^2 \theta \Big(r^2+a^2 + \frac{M a^2 \sin^2 \theta e^{-k/r^2}}{\rho^2}\Big), \nonumber\\
\xi_{t}\cdot \xi_{\phi} &=& g_{t \phi} =  \frac{M a \sin^2 \theta e^{-k/r^2}}{\rho^2},\nonumber\\
\xi_{\psi}\cdot \xi_{\psi} &=& g_{\psi \psi} = \cos^2 \theta \Big(r^2+b^2 + \frac{M b^2 \cos^2 \theta e^{-k/r^2}}{\rho^2} \Big), \nonumber\\
\xi_{t}\cdot \xi_{\psi} &=& g_{t \psi} = \frac{M b \cos^2 \theta e^{-k/r^2}}{\rho^2}, \nonumber\\
\xi_{\phi}\cdot \xi_{\psi} &=& g_{\phi \psi} =  \frac{M a b \sin^2 \theta \cos^2\theta e^{-k/r^2}}{\rho^2}
\end{eqnarray}
 The Killing vectors given in (\ref{killing}) can be used to give the physical interpretation of $a$ and $b$ of metric~(\ref{metric5}). Following \cite{Komar:1958wp}, we write the Komar integrals 
\begin{eqnarray}
&& \mathcal{M} = \frac{1}{4 \pi^2} \oint \xi_{(t)}^{a;b} d^3_{\Sigma_{a b}},    \nonumber\\
&& j_{(a)}=M a =-\frac{1}{4 \pi^2}  \oint \xi_{(\phi)}^{a;b} d^3_{\Sigma_{a b}}, \nonumber\\
&& j_{(b)}=M b =-\frac{1}{4 \pi^2}  \oint \xi_{(\psi)}^{a;b} d^3_{\Sigma_{a b}},
\end{eqnarray}
where the integrals $d^3_{\Sigma_{a b}}$ is defined as
\begin{equation}
d^3_{\Sigma_{a b}} =\frac{1}{3!} \sqrt{-g} \epsilon_{a b c d e} d x^{c}\wedge d x^{d}\wedge d x^{e},
\end{equation}
The semicolon ($;$) denotes the covariant differentiation, and $j_{(a)}$ and $j_{(b)}$ are the two specific angular momentum parameters in the $\phi$ and $\psi$ directions. We obtained the relation of mass $M$, total angular momentum $J_{(a)}$ and $J_{(b)}$ as given in \cite{Myers:1986un}
\begin{eqnarray}
\mathcal{M} = \frac{8}{3 \pi}M, \quad j_{(a)}=\frac{4}{\pi} J_{(a)}, \quad j_{(b)}=\frac{4}{\pi} J_{(b)}.
\end{eqnarray}
 In the four-dimensional case, a locally nonrotating observer includes a vector, which has a velocity orthogonal to the $t=const$ surface in the Kerr black hole background and vanished its angular momentum \cite{Bardeen:1970vja}. We define a local nonrotating observer in a similar way for the $5D$ rotating regular Myers-Perry spacetime. We write
 \begin{equation}
 u^{a}=u^{a}(r,\theta)=\alpha[\xi^{a}_{(t)} + \Omega_{(a)} \xi^{a}_{(\phi)} + \Omega_{(b)} \xi^{a}_{(\psi)}],
 \end{equation}
 where $u^{a}$ is the velocity vector of a nonrotating observer in $5D$ and $\alpha$ is a normalization constant, which is determined by $u^2=-1$. The orthogonality of $t=const$ surface implies $u^{r} = u^{\theta} =0$ and
 \begin{eqnarray}
 g_{t \phi} u^{t} + g_{\phi \phi} u^{\phi} + g_{\phi \psi} u^{\psi}=0, \nonumber\\ 
 g_{t \psi} u^{t} + g_{\psi \psi} u^{\psi} + g_{\psi \phi} u^{\phi}=0.
 \end{eqnarray}
  The simultaneous solution of these equations determines $u^{a} (r, \theta)$, and easily can be solved to get angular velocities
 \begin{eqnarray}
 \Omega_{(a)} &=&\frac{u^{\phi}}{u^t} = \frac{g_{t \psi} g_{\phi \psi} - g_{t \phi} g_{\psi \psi}}{g_{\phi \phi} g_{\psi \psi}-g_{\phi \psi}^2} = \frac{M a (r^2+b^2) e^{-k/r^2}}{\Delta \rho^2 + M (r^2+a^2)(r^2+b^2) e^{-k/r^2}}, \nonumber\\
 \Omega_{(b)} &=& \frac{u^{\psi}}{u^t} = \frac{g_{t \phi} g_{\phi \psi} - g_{t \psi} g_{\phi \phi}}{g_{\phi \phi} g_{\psi \psi}-g_{\phi \psi}^2} = \frac{M b (r^2+a^2) e^{-k/r^2}}{\Delta \rho^2 + M (r^2+a^2)(r^2+b^2) e^{-k/r^2}},
 \end{eqnarray}
 when either $b=0$ or $a=0$, one obtains
 \begin{eqnarray}
 \Omega_{(a)}=-\frac{g_{t \phi}}{g_{\phi \phi}} \quad \mbox{or} \quad  \Omega_{(b)}=-\frac{g_{t \psi}}{g_{\psi \psi}},
 \end{eqnarray}
  which at the horizon ($\Delta=0$) is given by
 \begin{eqnarray}
 \Omega_{(a)+}=\frac{a}{r^2_{+}+a^2}, \quad \quad \Omega_{(b)+}=\frac{b}{r^2_{+}+b^2}.
 \end{eqnarray}
\begin{figure}
    \begin{tabular}{c c }
        \includegraphics[scale=0.5]{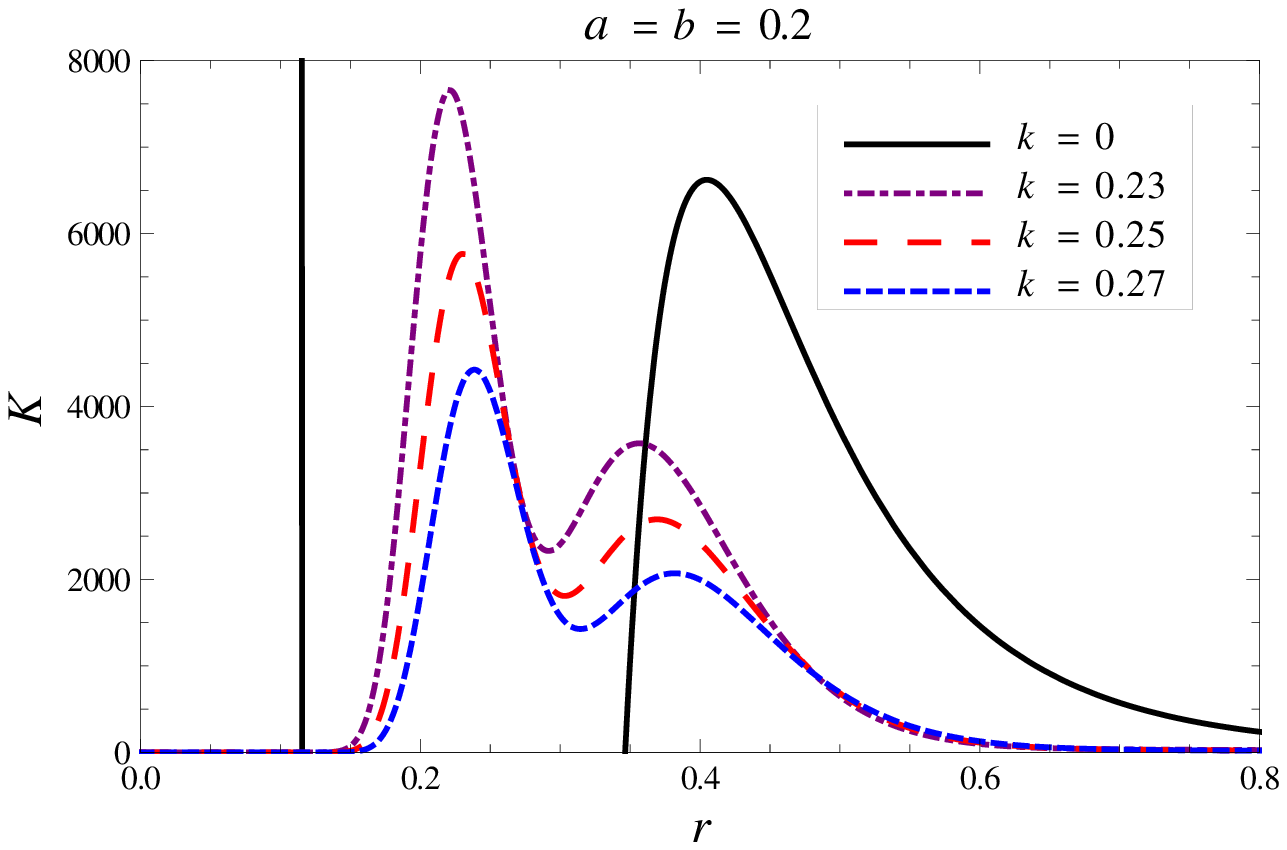}
        \includegraphics[scale=0.5]{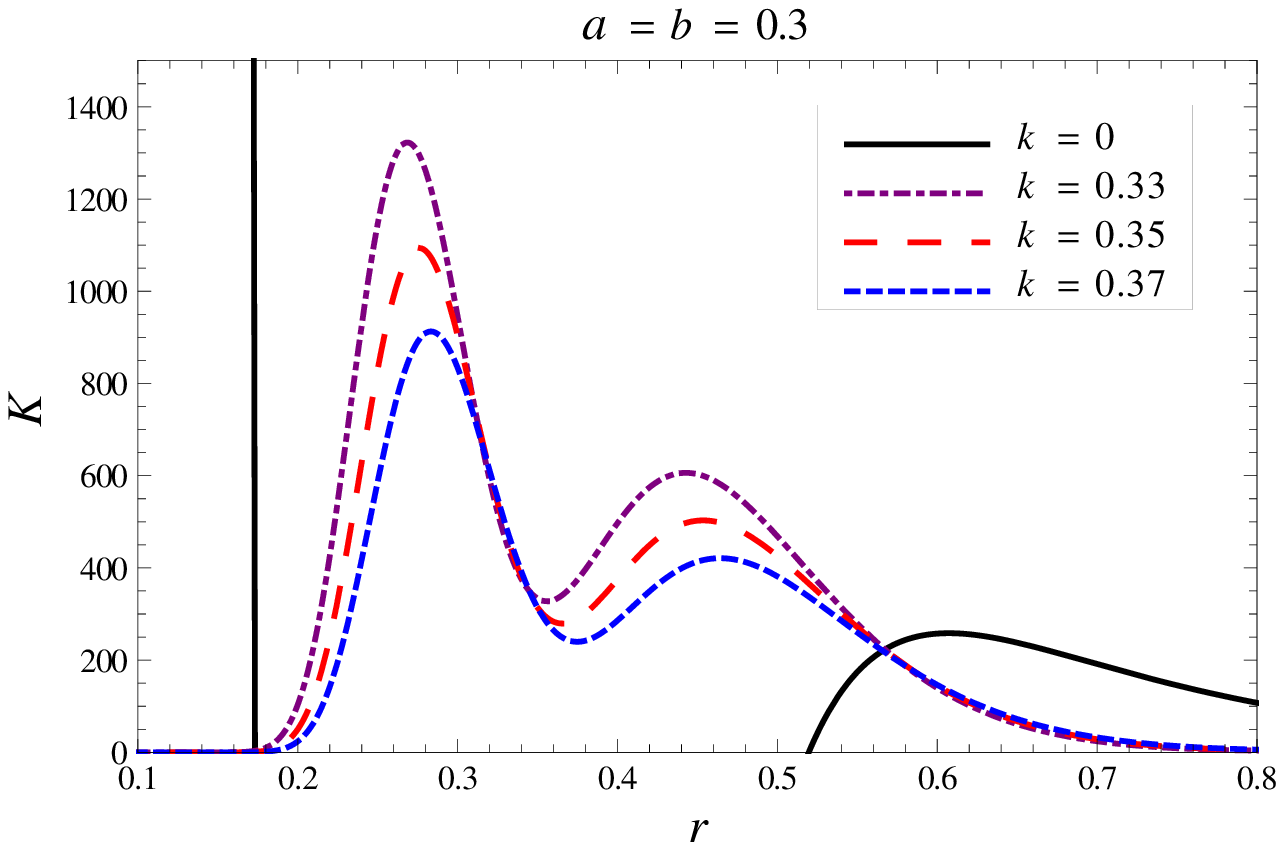}
    \end{tabular}
    \caption{\label{ks} Plots showing the regular behavior of Kretschmann scalar $K$ vs radius $r$ for the various values of the parameter $k$, and $\theta=\pi/2$. We have taken $M=1$.}
\end{figure}
 To check the regular behavior of the metric~(\ref{metric5}), we study the curvature invariants, i.e., Kretschmann scalar and Ricci scalar. If these curvature invariants are well behaved, then the obtained solution is regular. In the case of Myers-Perry black hole metric, these curvature invariants diverge at $r=0$. We calculated the Kretschmann scalar for $a=b$, and plotted with the radius $r$ for different values of the parameter $k$ (cf. Fig.~\ref{ks}). One can see that for every non-zero value of $k$, Kretschmann scalars are well defined. We have also examined other scalars for the metric~(\ref{metric5}), not reported here, and they are also regular everywhere. Hence, the obtained solution~(\ref{metric5}) is regular. 

\section{Horizons and ergoregions}\label{rrbh}
 The horizons of the $5D$ rotating regular black hole (\ref{metric5}) are the solution of $g^{rr}=\Delta=0$, i.e.,
\begin{eqnarray}\label{event}
r^4 + (a^2 + b^2) r^2 + a^2 b^2- M r^2 e^{-k/r^2} = 0.
\end{eqnarray}
For the given values of angular momenta $a$ and $b$ we have, for $k<k_c$, Eq.~(\ref{event}) admit two positive roots $(r_{\pm})$ corresponding to Cauchy $(r_-)$ and event $(r_+)$ horizons, where $k=k_c$ is a critical value. For equality $k=k_c$, the two horizons shrinks to one, this case corresponds to extremal black holes (Table~\ref{tb1} and Fig.~\ref{ehf}), while no black hole solutions when $k>k_c$.
  \begin{table}
 \begin{center}
 \caption{\label{tb1} Table for values of the outer horizon ($r_{+}$), inner horizn ($r_{-}$) and $\delta_r = r_{+}-r_{-}$. }
 \begin{tabular}{ l l l l  l l l }
 \hline \hline
 &\multicolumn{3}{c}{$a=0.2, b=0.3$}  & \multicolumn{3}{c}{$a=0.4, b=0.2$} \\
 \hline  
$k$ & $r_{+}$ & $r_{-}$ & $\delta_r$ & $r_{+}$ & $r_{-}$ & $\delta_r$ \\
  \hline
0  \,\, & 0.93050  & 0.06448   & 0.86602     \,\, & 0.88989  & 0.08989  & 0.8   \\
0.05  \,\, & 0.89740  & 0.19362   & 0.70377      \,\, & 0.85116  & 0.22617  & 0.62499 \\
0.15  \,\, & 0.81347  & 0.34224   & 0.47123      \,\, & 0.74124  & 0.40696  & 0.33428 \\
$k_c$  \,\,& 0.59783 & 0.59783   & \,\,\,\,\,0      \,\, & 0.58546  & 0.58546  & \,\,\,\,\,0      \\
 \hline \hline
  \end{tabular}
 \end{center}
\end{table}

 \begin{figure}[h]
 \begin{center}
 \begin{tabular}{c c c}
        \includegraphics[scale=0.4]{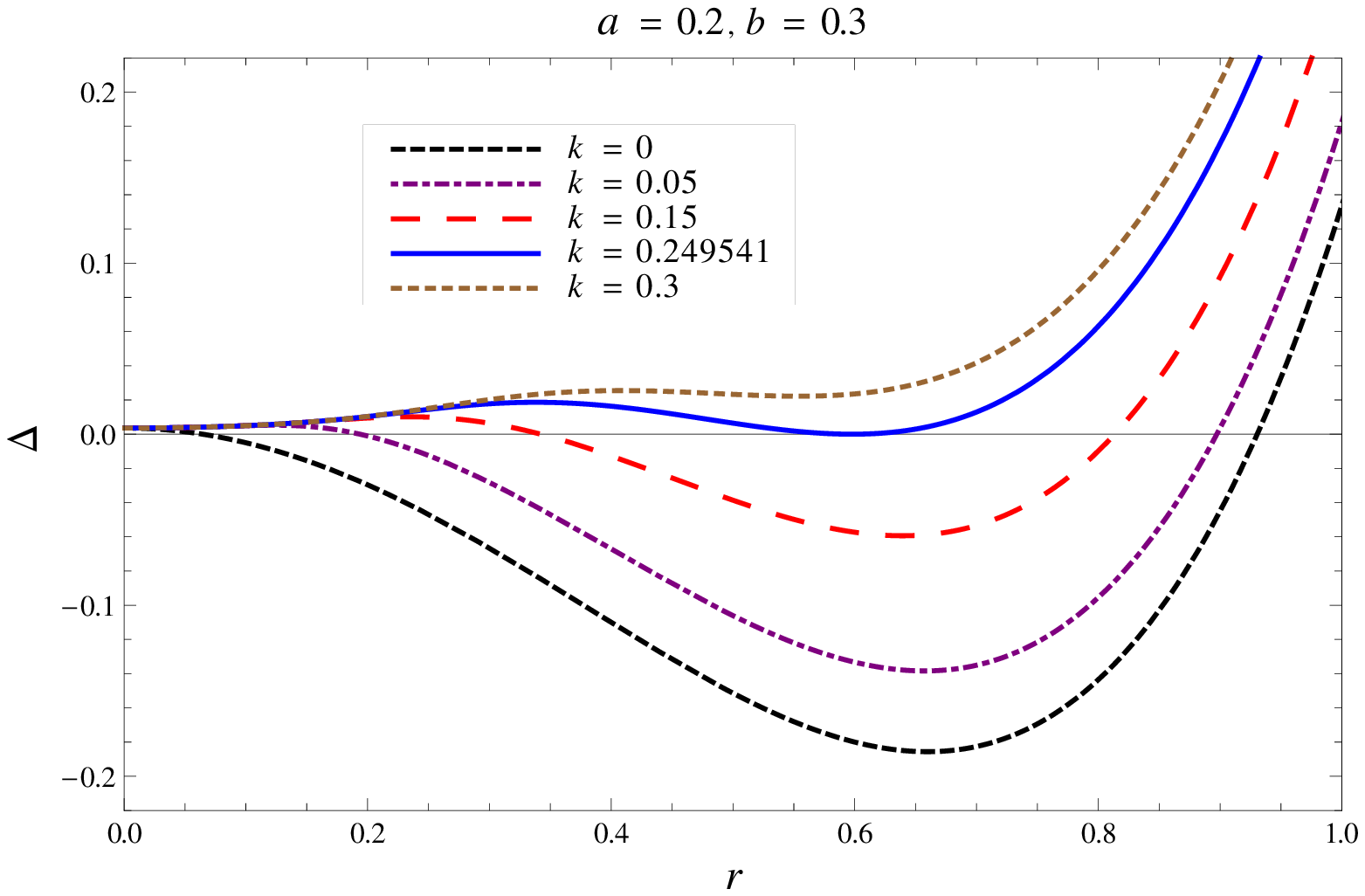}
        \includegraphics[scale=0.4]{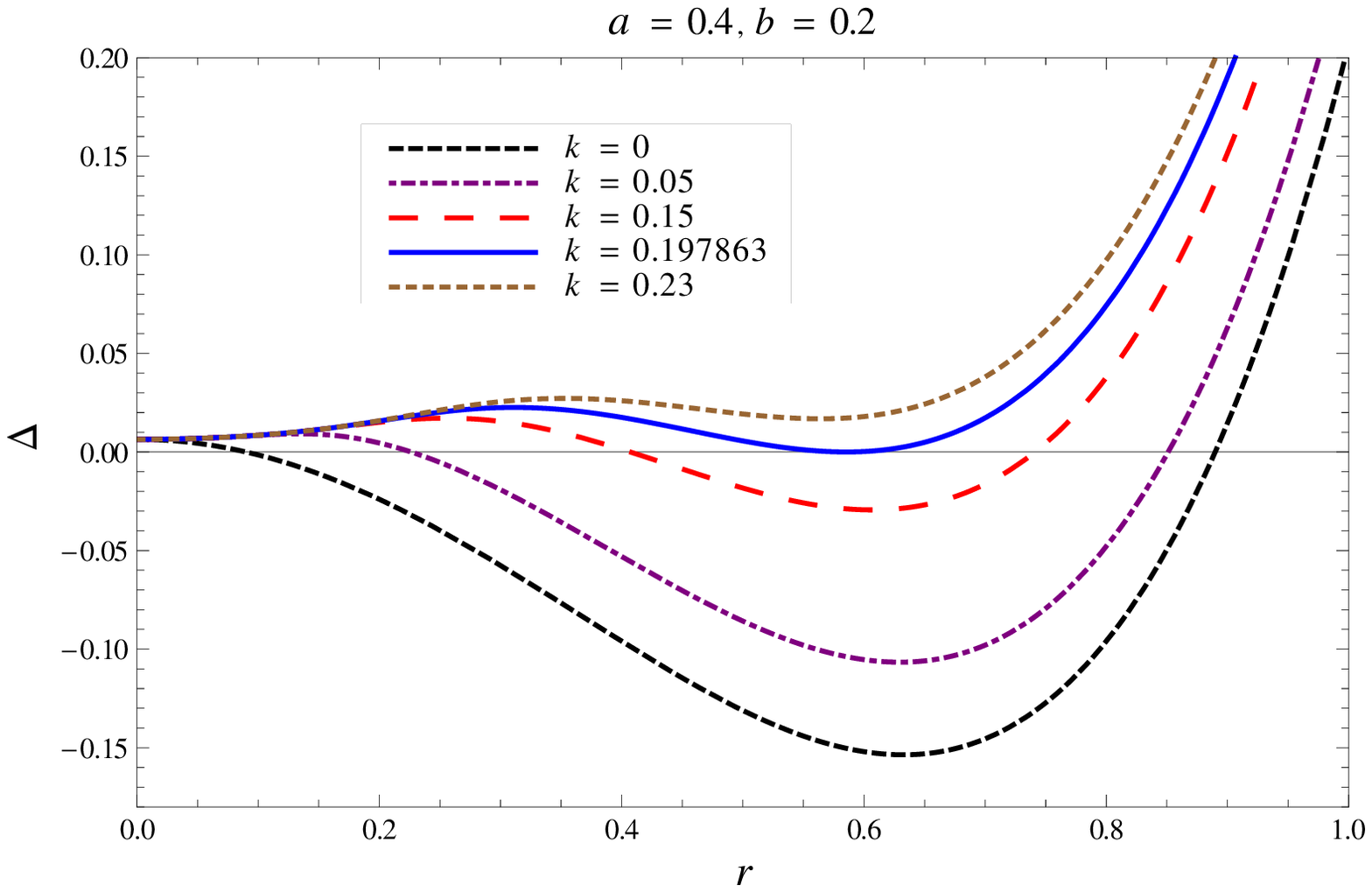}
    \end{tabular}  
     \end{center}   
    \caption{Plots showing the behavior of $\Delta$ vs radius $r$ for different values of parameter $k$.}\label{ehf}
\end{figure}

 \begin{figure}
  \begin{center}
 \begin{tabular}{c c}
\includegraphics[scale=0.4]{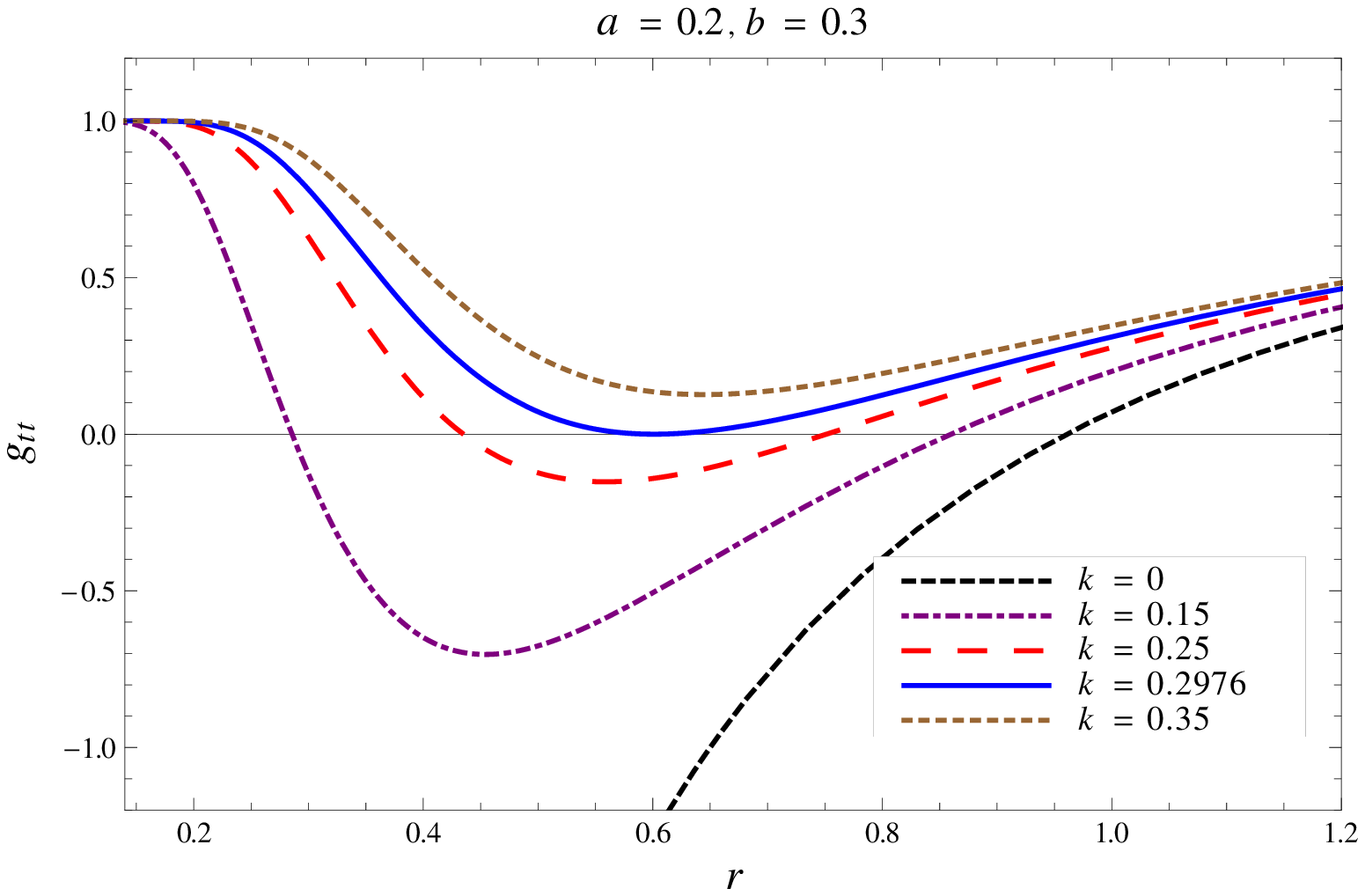}
\includegraphics[scale=0.4]{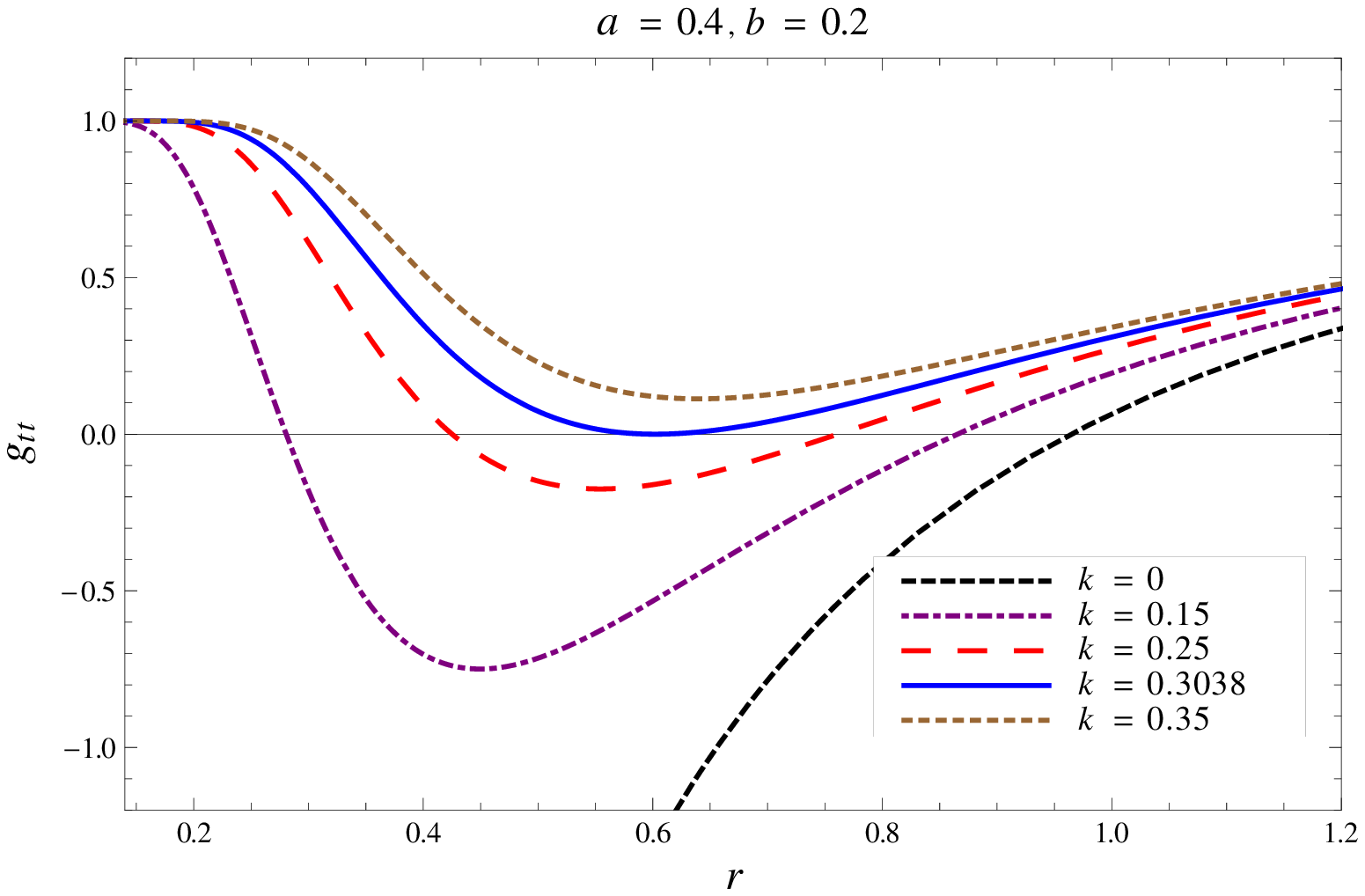}
 \end{tabular}
 \end{center} 
\caption{Plots showing the behavior of -$g_{tt}$ vs radius $r$ for the different values of parameter $k$.}\label{slsf}
\end{figure}

 The static limit surface is a surface of infinite red-shift and time-translation Killing vector becomes null there, which gives $g_{tt}=0$, i.e.,
  \begin{eqnarray}\label{static}
 r^2 + a^2 \cos^2 \theta+b^2 \sin^2 \theta - M e^{-k/r^2} = 0.
\end{eqnarray}
From Eq.~(\ref{static}), it is clear that the static limit surface depends on the values of the rotation parameter $a$ and $b$, free parameter $k$ and angle $\theta$. The static limit surface has the oblate shape and its size shrinks from equator to pole and at the pole, it coincides with the event horizon. In Fig.~\ref{slsf}, we plot the behavior of the  $g_{tt}$ with $r$ and found that the static limit surface also has a critical value of the free parameter $k$. 
\begin{figure*}
\begin{tabular}{c c c c}
\includegraphics[scale=0.4]{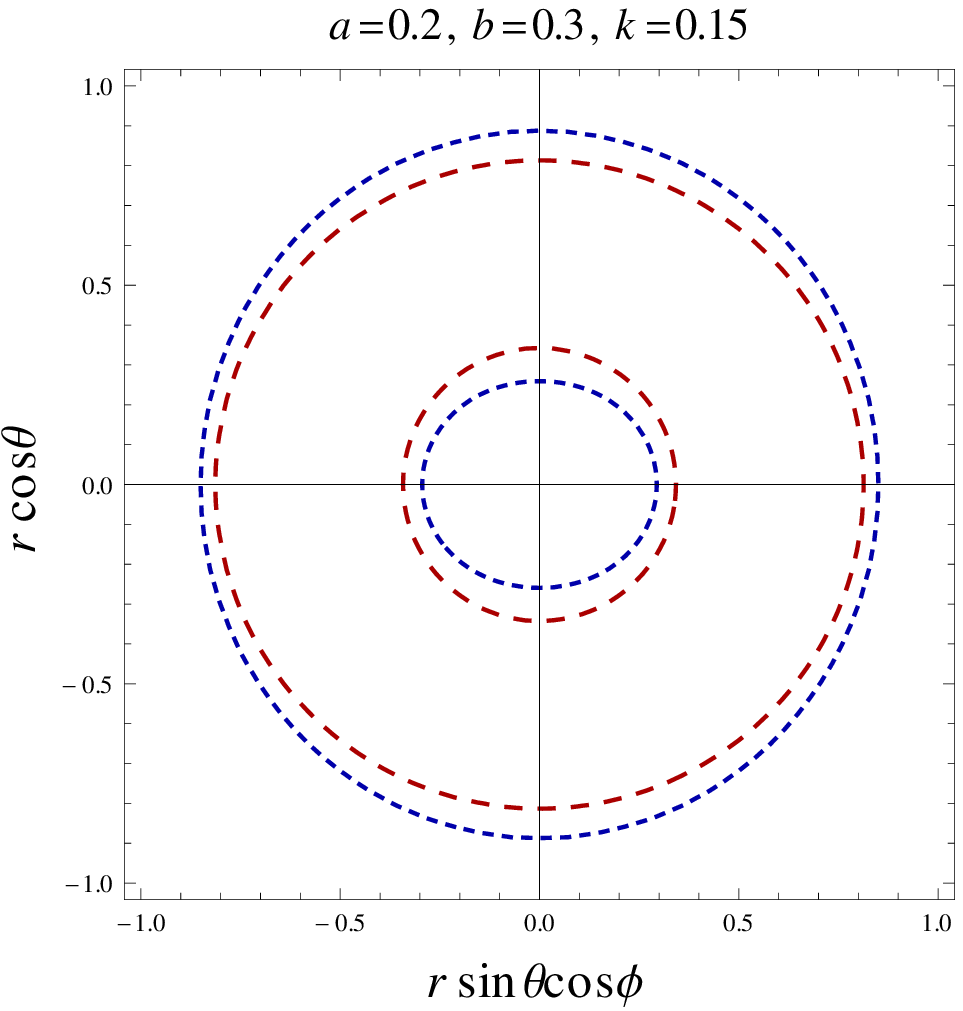}&
\includegraphics[scale=0.4]{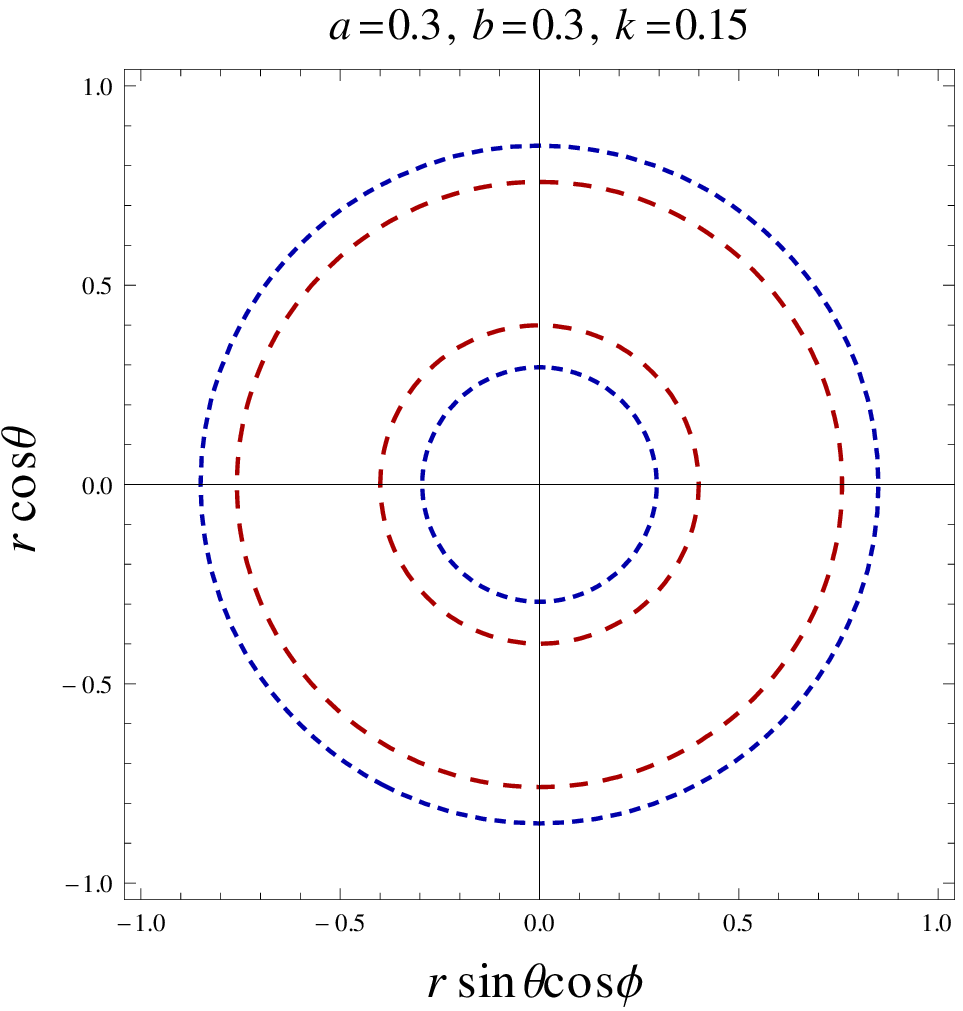}&
\includegraphics[scale=0.4]{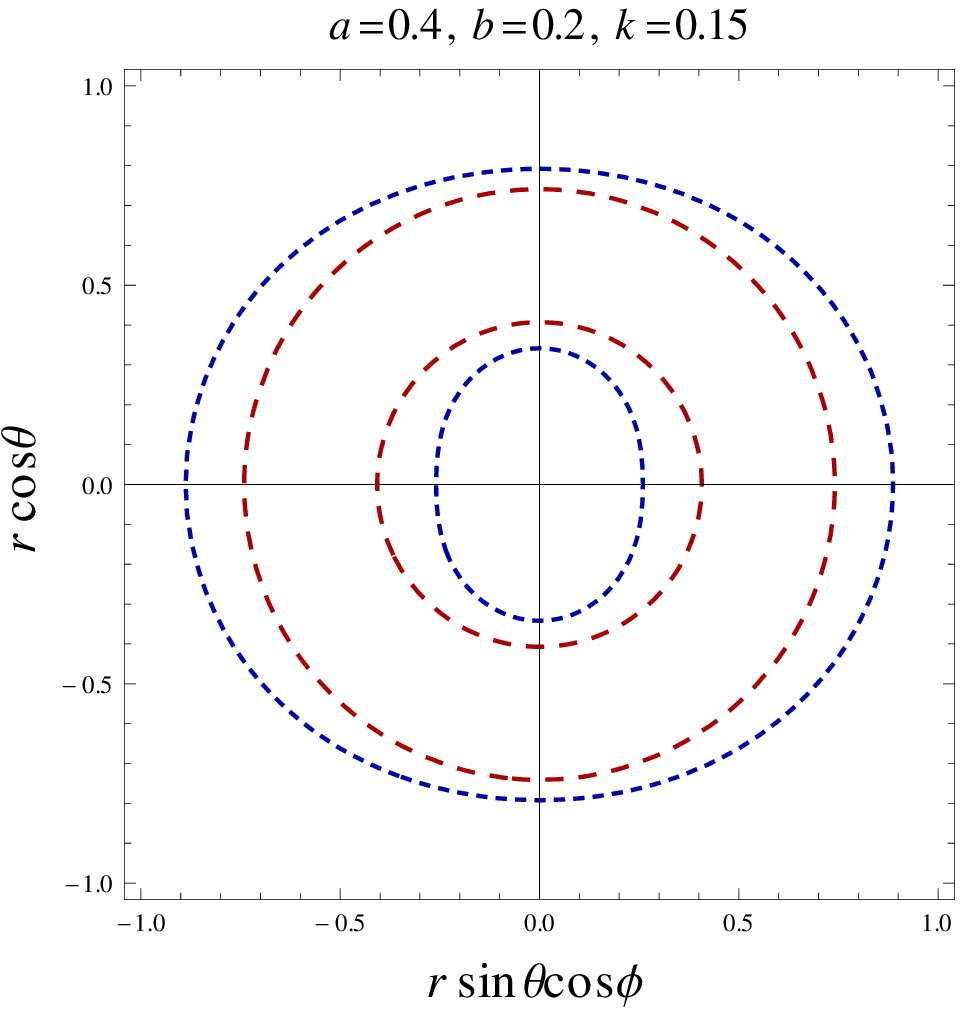}\\
\includegraphics[scale=0.4]{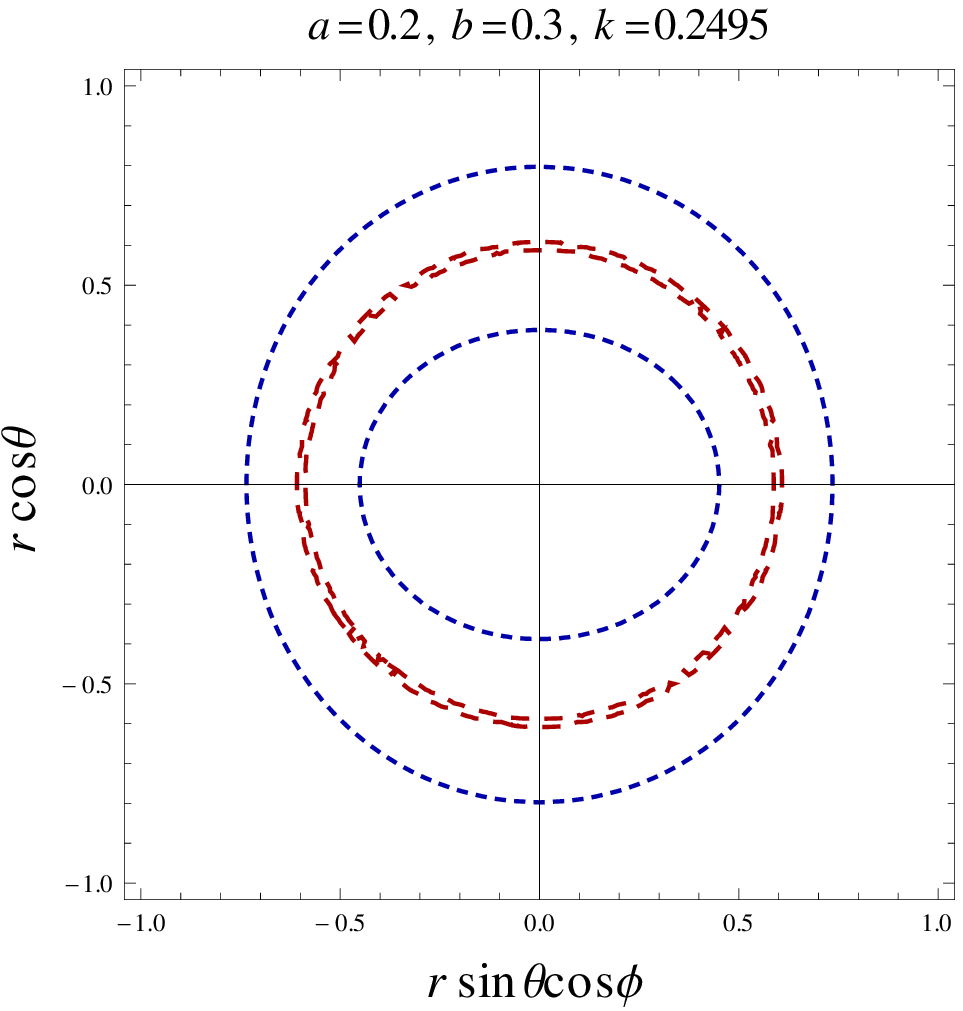}&
\includegraphics[scale=0.4]{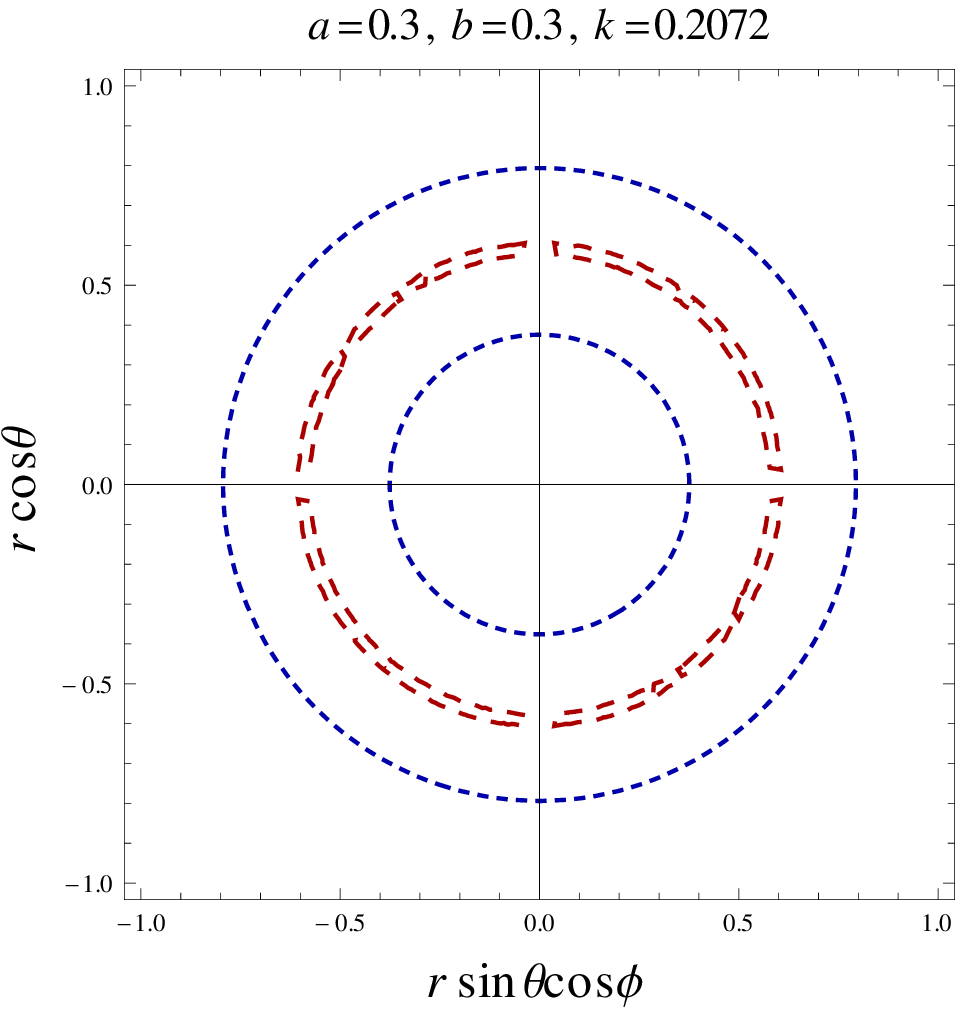}&
\includegraphics[scale=0.4]{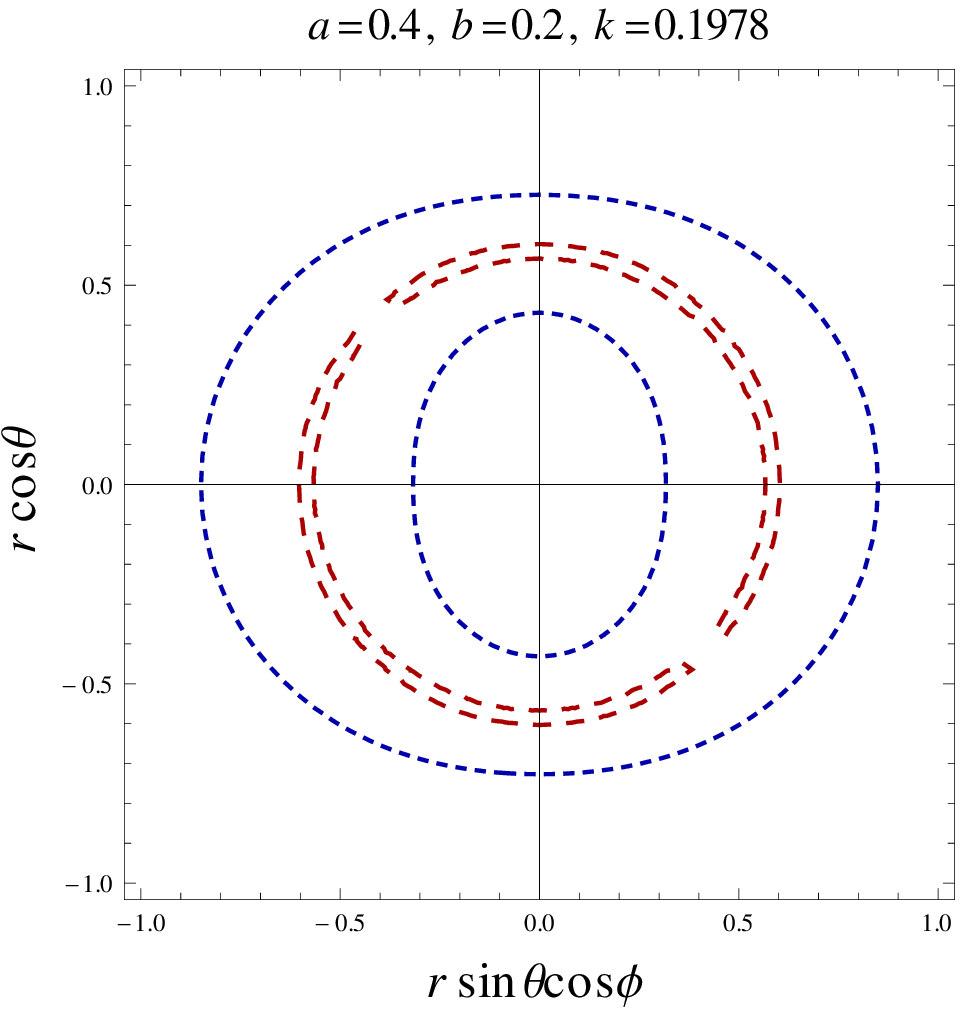}\\
\includegraphics[scale=0.4]{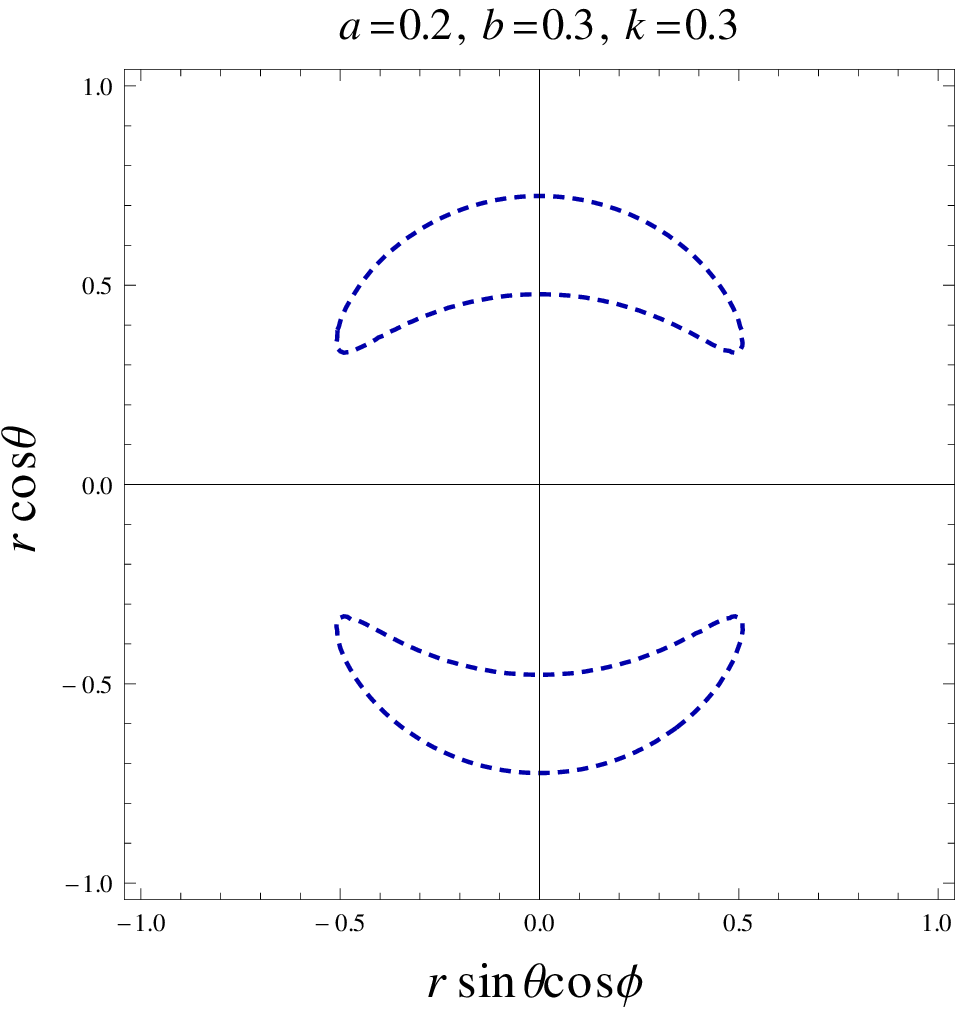}&
\includegraphics[scale=0.4]{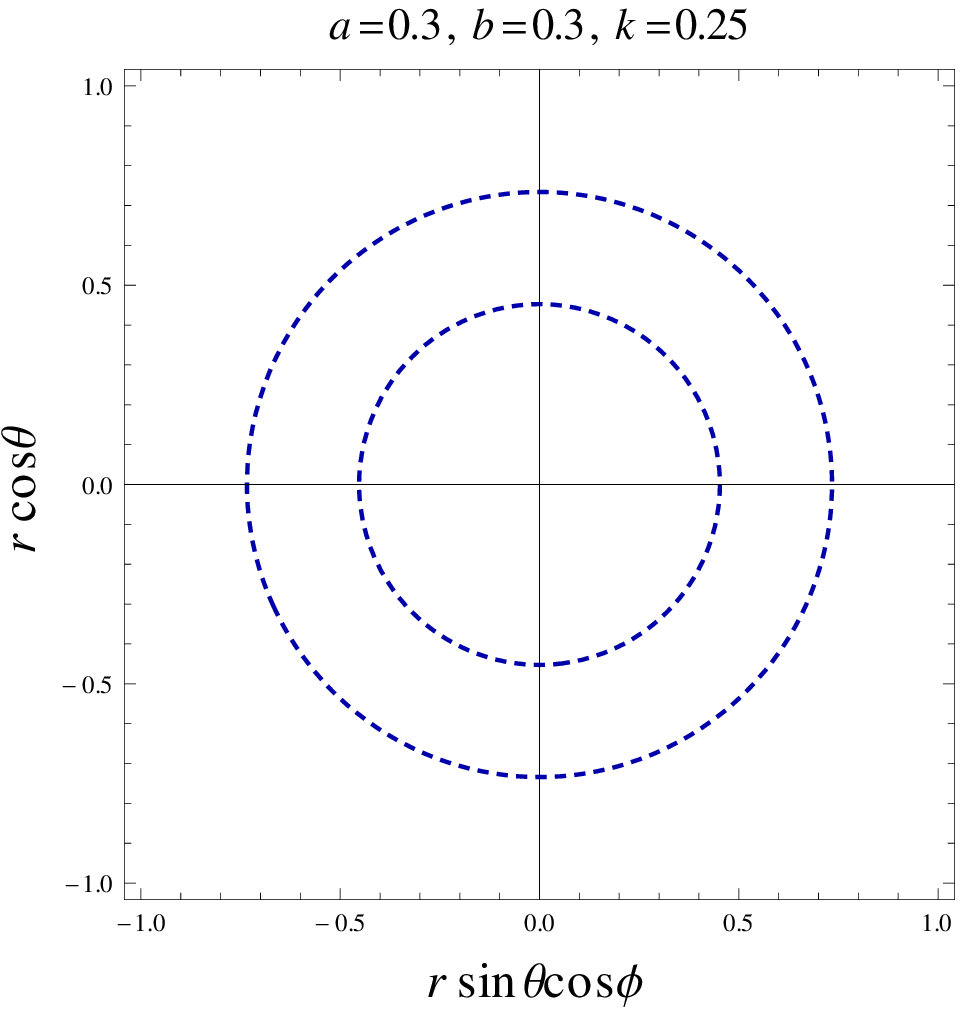}&
\includegraphics[scale=0.4]{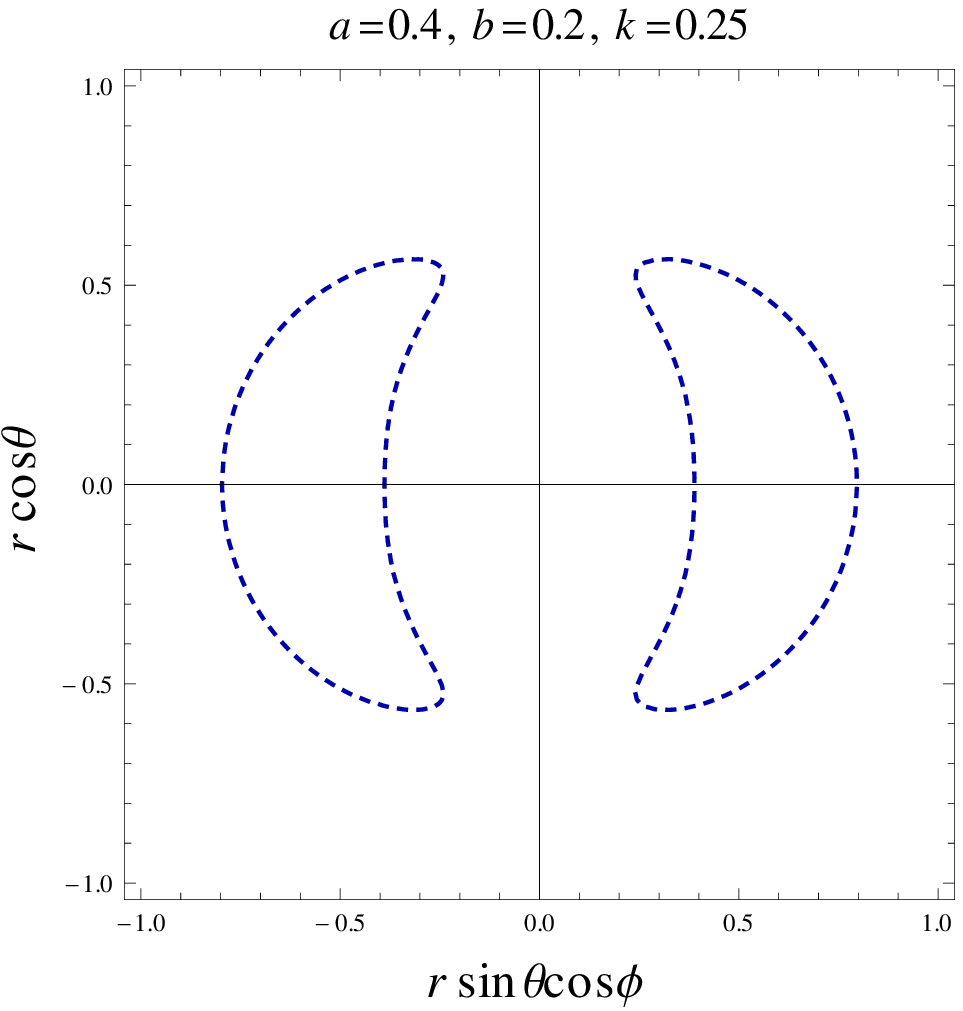}\\
\end{tabular}
\caption{Plots showing the behavior of ergo-region with rotation parameter $a,b$ and free parameter $k$.}\label{ergo}
\end{figure*}
\\ The ergoregion is defined as the region between the static limit surface and the event horizon. It is important from the astrophysical points of view because some important phenomena take place in the region like Penrose Process \cite{Penrose:1971uk}, i.e., the extraction of rotational energy. A particle or object is free to enter and leave the ergoregion, and an object cannot remain static in the ergoregion, but it moves with the rotation of the black hole, which is known as a frame-dragging effect. The ergoregion corresponding in parameter ($a,b,k$) are plotted in Fig.~\ref{ergo} (see also Table~\ref{tb2}).
\begin{table}
 \begin{center}
 \caption{\label{tb2} Table for radius of event horizon ($r_{+}$), static limit surface ($r_{+}^{SLS}$) and $\delta=r_{+}^{SLS}-r_{+}$ for different values of parameter $k$.}
 \begin{tabular}{l l l l  l l l l l l }
 \hline 
 &\multicolumn{3}{c}{$a=0.2, b=0.3$}  & \multicolumn{3}{c}{$a=0.4, b=0.2$} \\
 \hline  
$k$ & $r_{+}$ & $r_{+}^{SLS}$ & $\delta$ & $r_{+}$ & $r_{+}^{SLS}$ & $\delta$    \\
  \hline
0      & 0.93050  & 0.96046  & 0.02996  & 0.88989  & 0.96436  & 0.07446   \\
0.05   & 0.89740  & 0.93082  & 0.03341  & 0.85116  & 0.93510  & 0.08394   \\
0.10   & 0.85934  & 0.89765  & 0.03831  & 0.80428  & 0.90247  & 0.09819 \\
0.15   & 0.81347  & 0.85949  & 0.04602  & 0.74124  & 0.86508  & 0.12384  \\
\hline \hline
  \end{tabular}
 \end{center}
\end{table}

\section{Black hole shadow}\label{pmbh}
The shape and size of the black hole shadow depend on the spacetime geometry. The shadow of a black hole is an apparent geometry of the event horizon. A black hole casts a shadow if it is situated between the light source like Quasars and the observer.
Recently, it becomes the field of intensive research, both in four dimensions
\cite{Akiyama:2019eap,Cunha:2015yba,Bohn:2014xxa,Yumoto:2012kz,Perlick:2018iye,Synge:1966okc,bardeen2,Bambi:2008jg,Abdujabbarov:2016hnw,Amir:2016cen,Amarilla:2011fxx,Grenzebach:2014fha,Amarilla:2010zq,Wei:2013kza,Akiyama:2019fyp,Akiyama:2019cqa,Younsi:2016azx,Abdujabbarov:2015xqa,Atamurotov:2013sca,Amarilla:2013sj}
 and in higher dimensions \cite{Hertog:2019hfb,Amir:2017slq,Papnoi:2014aaa}.
When the incoming photons emitted by a light source, come near to a black hole, then they follow the three types of trajectories, i.e., infalling orbits, scattering orbits, and unstable circular orbits. The unstable circular orbit helps us to analyze the shape and size of the black hole shadow. First, we analyze the geodesic structure of a particle moving around $5D$ rotating regular Myers-Perry black holes.   
The Lagrangian reads
\begin{equation} \label{canonical1}
\mathcal{L}= \frac{1}{2} g_{a b}u^{a}u^{b},
\end{equation}
where $u^{a}$ is the 5-velocity of the particle. The canonical momenta are calculated as
\begin{widetext}
\begin{eqnarray}\label{canonical}
p_{t}&=&\Big(-1+\frac{ M e^{-k/r^2}}{\rho^2}\Big)u^{t}+\Big(\frac{M a \sin^2 \theta e^{-k/r^2}}{\rho^2}\Big)u^{\phi}  + \Big(\frac{  M b \cos^2 \theta e^{-k/r^2}}{\rho^2}\Big)u^{\psi},\nonumber\\
p_{\phi}&=&\Big(\frac{M a \sin^2 \theta e^{-k/r^2}}{\rho^2}\Big)u^{t} + \Big(r^2+a^2+\frac{M a^2 \sin^2 \theta}{\rho^2}\Big)\sin^2\theta u^{\phi}  + \Big(\frac{M a b \sin^2\theta \cos^2\theta e^{-k/r^2}}{\rho^2}\Big)u^{\psi},\nonumber\\
p_{\psi}&=& \Big(\frac{M b \cos^2 \theta e^{-k/r^2}}{\rho^2}\Big)u^{t}+\Big(\frac{  M a b \sin^2 \theta \cos^2 \theta e^{-k/r^2}}{\rho^2}\Big)u^{\phi}+\Big(r^2+b^2+\frac{  M b^2 \cos^2 \theta e^{-k/r^2}}{\rho^2}\Big)\cos^2 \theta u^{\psi}, \nonumber\\
p_{r}&=& \frac{\rho^2 r^2}{\Delta}u^{r},  \;\;\;\;\;  p_{\theta}= \rho^2 u^{\theta}.
\end{eqnarray}
\end{widetext}
Metric~(\ref{metric5}), has three conserved quantities associated with energy $E$, and angular momenta $L_{\phi}$ and $L_{\psi}$, and we have $p_{t}=-E$,  $p_{\phi}=L_{\phi}$ and $p_{\psi}=L_{\psi}$. Solving the first three equations from Eq.~(\ref{canonical}), we get
\begin{widetext}
\begin{eqnarray}\label{tphi}
u^{t}&=&\frac{1}{\rho^2 \Delta}\Big(\Big(\rho^2  \Delta + M (r^2+a^2)(r^2+b^2) e^{-k/r^2}\Big) E + M a (r^2+b^2)e^{-k/r^2} L_{\phi}  +  M b (r^2+a^2)e^{-k/r^2} L_{\psi} \Big),\nonumber\\
u^{\phi}&=&\frac{1}{\rho^2  \Delta} \Big(- M a(r^2+b^2) e^{-k/r^2} E + \Big(\frac{\rho^2  \Delta - M a^2 (r^2 +b^2) \sin^2 \theta e^{-k/r^2}}{(r^2+a^2) \sin^2 \theta}\Big) L_{\phi}  -  M a b e^{-k/r^2} L_{\psi}  \Big), \nonumber\\
u^{\psi}&=& \frac{1}{\rho^2  \Delta} \Big(- M b(r^2+a^2) e^{-k/r^2} E + \Big(\frac{\rho^2  \Delta - M b^2 (r^2 + a^2) \cos^2 \theta e^{-k/r^2}}{(r^2+b^2) \cos^2 \theta}\Big) L_{\psi} - M a b e^{-k/r^2} L_{\phi}  \Big).
\end{eqnarray}
\end{widetext}
We use the Hamilton-Jacobi equation to separate the radial and angular part of the equation of motion \cite{Frolov:2003en}. The most general Hamilton-Jacobi equation reads 
\begin{eqnarray}\label{hamilton}
-\frac{\partial S}{\partial \lambda}=\frac{1}{2}g^{a b}\frac{\partial S}{\partial x^{a}}\frac{\partial S}{\partial x^{b}}
\end{eqnarray}
where $\lambda$ is an affine parameter and $S$ is the Jacobi action, which for the $5D$ case reads
\begin{eqnarray}\label{action}
S=\frac{1}{2}m^2 \lambda - E t + L_{\phi} \phi + L_{\psi} \psi + S_{\theta} (\theta) + S_{r}(r),
\end{eqnarray}
where $m$, $E$ and $L$, respectively the particle mass, energy and angular momentum. $S(r)$ and $S(\theta)$ are function of $r$ and $\theta$.
Since we are calculating the equations of motion for the photon, so the mass of the particle is set to be zero ($m=0$). Combining Eq.~(\ref{hamilton}) and Eq.~(\ref{action}), we obtain
\begin{eqnarray}\label{stheta}
\Big(\frac{\partial S_{\theta}}{\partial \theta}\Big)^2 - E^2 ( a^2 \cos^2 \theta + b^2 \sin^2 \theta )  + \frac{L_{\phi}^2}{\sin^2 \theta}+ \frac{L_{\psi}^2}{\cos^2 \theta}-\mathcal{K} = 0,
\end{eqnarray}
and 
\begin{eqnarray}\label{sr}
&& \Delta \Big(\frac{\partial S_r}{\partial r}\Big)^2 - E^2 r^2  -\frac{  M (r^2+a^2)(r^2 +b^2) e^{-k/r^2}}{\Delta }\Big(E+\frac{a L_{\phi}}{r^2+a^2}+\frac{b L_{\psi}}{r^2+b^2}\Big)^2 \nonumber\\&& - (a^2-b^2)\Big(\frac{L_{\phi}^2}{r^2+a^2}-\frac{L_{\psi}^2}{r^2+b^2}\Big) +\mathcal{K} = 0,
\end{eqnarray}
where $\mathcal{K}$ is a Carter separable constant \cite{Carter:1968rr}. We can write Eqs.~(\ref{stheta}) and (\ref{sr}) into the following form
\begin{eqnarray}
\frac{\partial S_{\theta}}{\partial \theta} =\pm \sqrt{\Theta} \quad \mbox{and} \quad \frac{\partial S_r}{\partial r} =\pm \sqrt{\mathcal{R}},
\end{eqnarray}
where $\Theta$ and $\mathcal{R}$ are given by
\begin{eqnarray}\label{theta}
 \Theta &=& E^2 (a^2 \cos^2 \theta + b^2 \sin^2 \theta) - \frac{L_{\phi}^2}{\sin^2 \theta} - \frac{L_{\psi}^2}{\cos^2 \theta} + \mathcal{K}, \nonumber\\
\mathcal{R} &=& \Delta  \Big(E^2 r^2 + (a^2-b^2)\Big(\frac{L_{\phi}^2}{r^2+a^2}-\frac{L_{\psi}^2}{r^2+b^2}\Big) -\mathcal{K}\Big) \nonumber\\&& +  M (r^2+a^2)(r^2+b^2)e^{-k/r^2}\Big(E+\frac{a L_{\phi}}{r^2+a^2}+\frac{b L_{\psi}}{r^2+b^2}\Big)^2. \nonumber\\
\end{eqnarray}
Eq.~(\ref{theta}) give the angular and radial part of the equation of motion of the photon. 
The effective potential has much importance in the study of the circular orbit of the photon around the black hole, which is given by
\begin{equation}
\frac{1}{2} (u^r)^2 + V_{eff} = 0,
\end{equation}
where 
\begin{eqnarray}\label{eff}
V_{eff} &=& -\frac{1}{2 \rho^4}\Big(\Delta  \Big(E^2 r^2 + (a^2-b^2)\Big(\frac{L_{\phi}^2}{r^2+a^2}-\frac{L_{\psi}^2}{r^2+b^2}\Big)  -\mathcal{K}\Big) \nonumber\\&& +  M (r^2+a^2)(r^2+b^2)e^{-k/r^2}   \Big(E+\frac{a L_{\phi}}{r^2+a^2}+\frac{b L_{\psi}}{r^2+b^2}\Big)^2\Big).
\end{eqnarray}
Note, we study the geometry of photons near the $5D$ rotating regular Myers-Perry black holes. Let us introduce the impact parameters for $5D$ case as
\begin{eqnarray}
\xi_1 = \frac{L_{\phi}}{E}, \quad \quad  \xi_2 = \frac{L_{\psi}}{E}, \quad \quad \eta = \frac{\mathcal{K}}{E^2}.
\end{eqnarray}
Take $b=a$ for simplicity, and write the $\mathcal{R}$ in terms of new parameters
\begin{equation}\label{R1}
\mathcal{R} = \Delta  (r^2 -\eta)+ M (r^2+a^2)^2 e^{-k/r^2} \Big(1+\frac{a}{r^2+a^2}(\xi_1+\xi_2)\Big)^2.
\end{equation}
The black hole shadow is determined by the unstable circular photon orbits which satisfy
\begin{equation}\label{circular}
 \mathcal{R} =  \frac{d \mathcal{R}}{d r }= 0.
\end{equation}
The impact parameters $\xi$ and $\eta$ determine the contour of the shadow for the photon orbits around the black hole. The value of $\eta$ and $\xi$ can be obtained from Eqs.~(\ref{R1}) and (\ref{circular}) as
\begin{widetext}
 \begin{eqnarray}\label{eta}
&& \eta=   \frac{r^2}{a\Big(r^4+e^{k/r^2}(a^2+r^2)((a^2+r^2)k-2 r^4)\Big)^2}\Big(a e^{2k/r^2}(r^2+a^2)^2((a^2+r^2)k -a^2 r^2  \nonumber\\&& -3 r^4)((a^2+r^2)k-2 r^4) +2 r^2 \sqrt{-e^{k/r^2} r^4 (a^2+r^2)((a^2+r^2)k  -2 r^4)} (a r^2-a e^{k/r^2}(a^2+r^2)^2)\nonumber\\&& + a e^{k/r^2} r^4(a^2+r^2)(3(a^2+r^2)k-5 r^4+a^2 r^2)\Big),\nonumber\\
&& \xi =  \frac{1}{ a\Big( r^4+e^{k/r^4}(a^2+r^2)((a^2+r^2)k-2 r^4)\Big)}\Big(- a^2 r^4 -  e^{k/r^2}(r^2+a^2)^2  ((a^2+r^2)k  -r^4) \nonumber\\ && +\sqrt{-e^{k/r^2}r^4(a^2+r^2)((a^2+r^2)k-2 r^4)}(- r^2+ e^{k/r^2}(a^2+r^2)^2)\Big) =  \xi_1+\xi_2.
\end{eqnarray}
\end{widetext}
Now, we consider the case, when $\theta=\pi/2$, $L_{\psi}=0$, which implies $\xi_2=0$, so from Eq.~(\ref{eta}), we obtain 
\begin{eqnarray}
\xi_1 &=&  \frac{1}{ a \Big( r^4+e^{k/r^4}(a^2+r^2)((a^2+r^2)k-2 r^4)\Big)}  \Big(- a^2 r^4 - e^{k/r^2}(r^2+a^2)^2((a^2+r^2)k-r^4)\nonumber\\ && + (- r^2+ e^{k/r^2}(a^2+r^2)^2)   \sqrt{-e^{k/r^2}r^4(a^2+r^2)((a^2+r^2)k-2 r^4)} \Big).
\end{eqnarray}
\begin{figure*}
\includegraphics[scale=0.42]{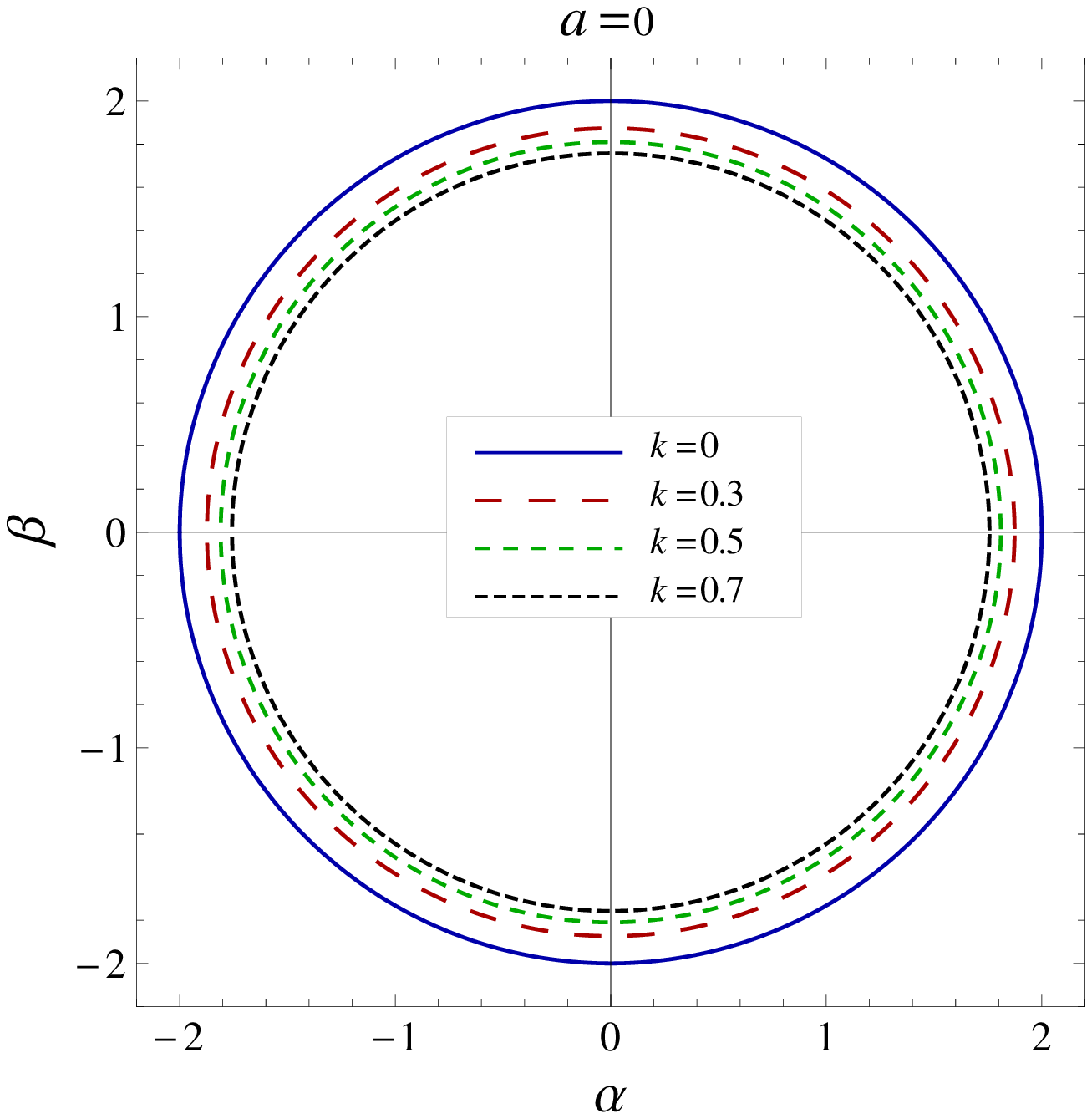}
\includegraphics[scale=0.55]{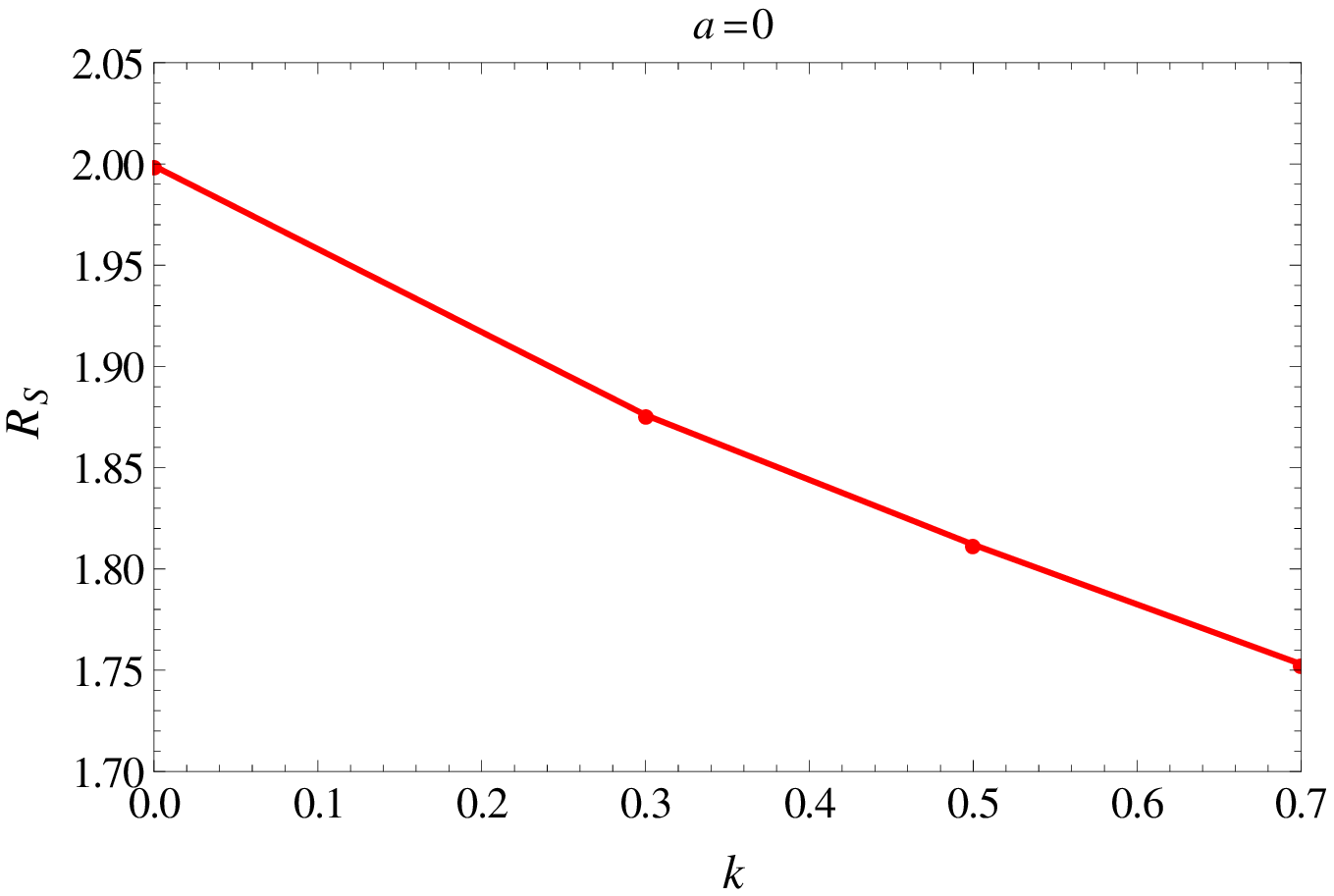}
\caption{\label{fig5} Plot showing the shapes of the black shadow for non-rotating regular Myers-Perry black hole with different values of the free parameter $k$.}
\end{figure*}
Clearly, $0\leq r \leq \infty$.
Now, we consider the case when $a=0$, then the value of $\eta$ take the following form
\begin{widetext}
\begin{eqnarray}\label{eta0}
\eta &=& \frac{4}{2-e^{-k/2}}.
\end{eqnarray}
\end{widetext}

\begin{figure*}
\includegraphics[scale=0.32]{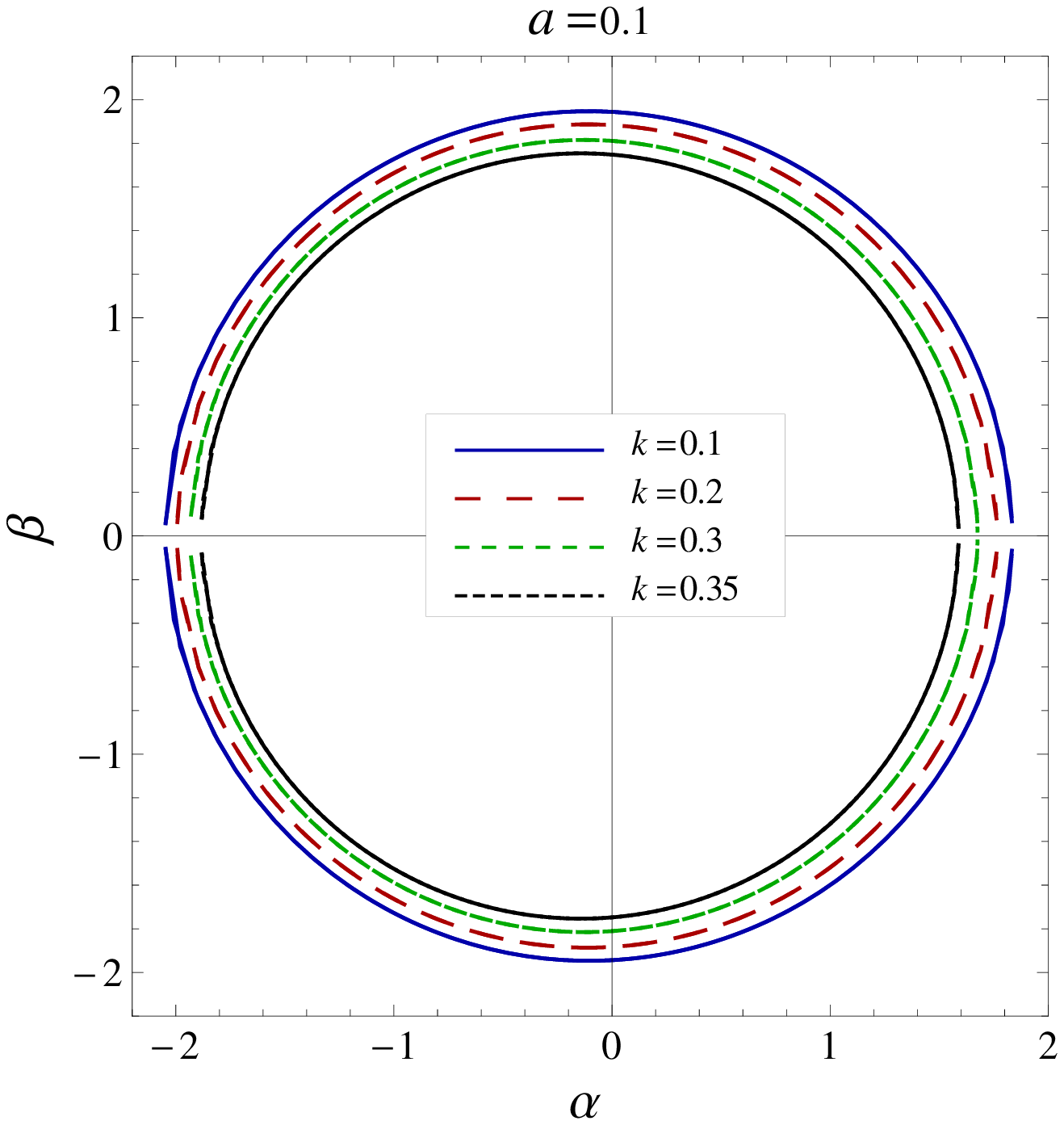}
\includegraphics[scale=0.34]{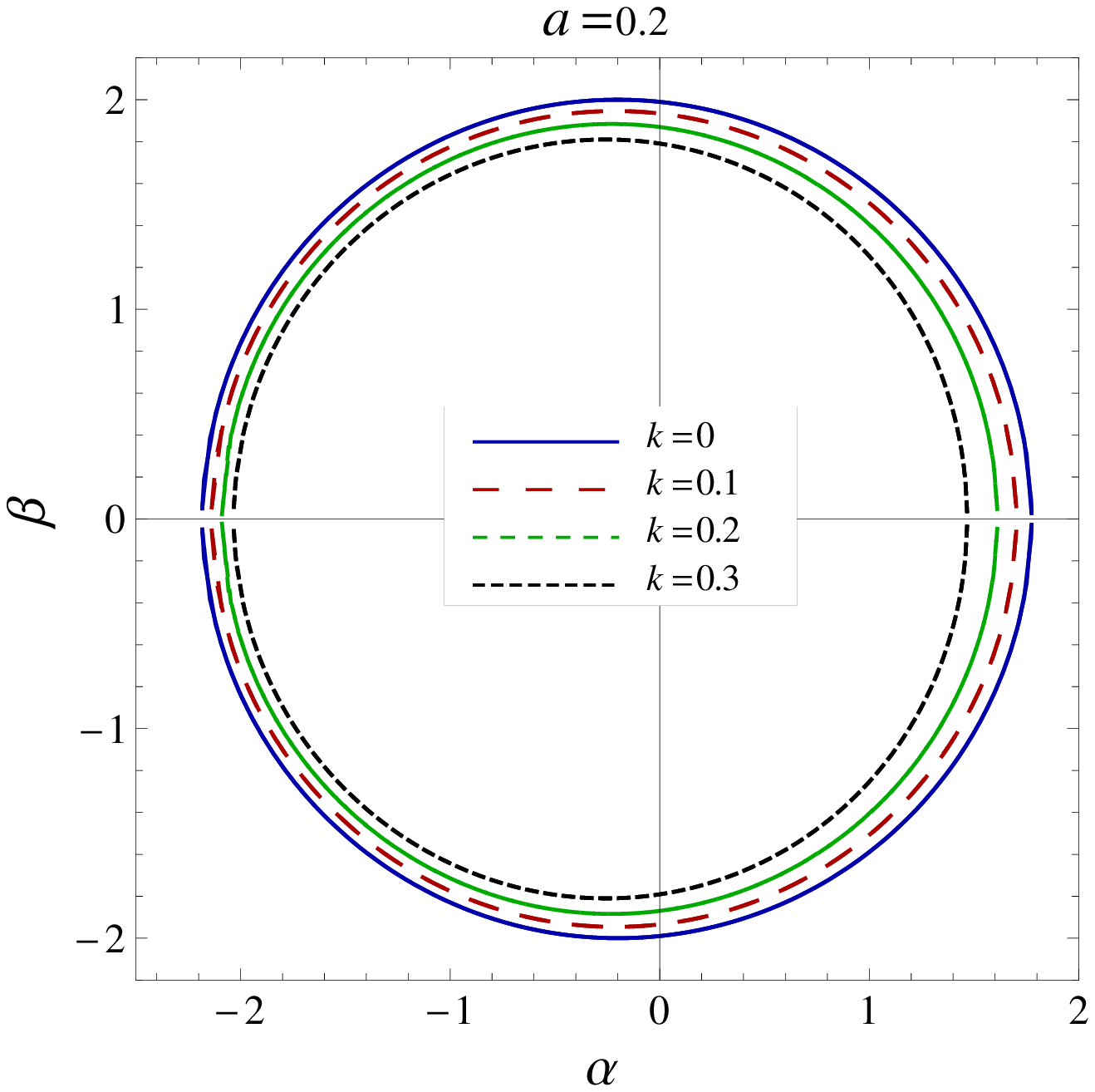}
\includegraphics[scale=0.34]{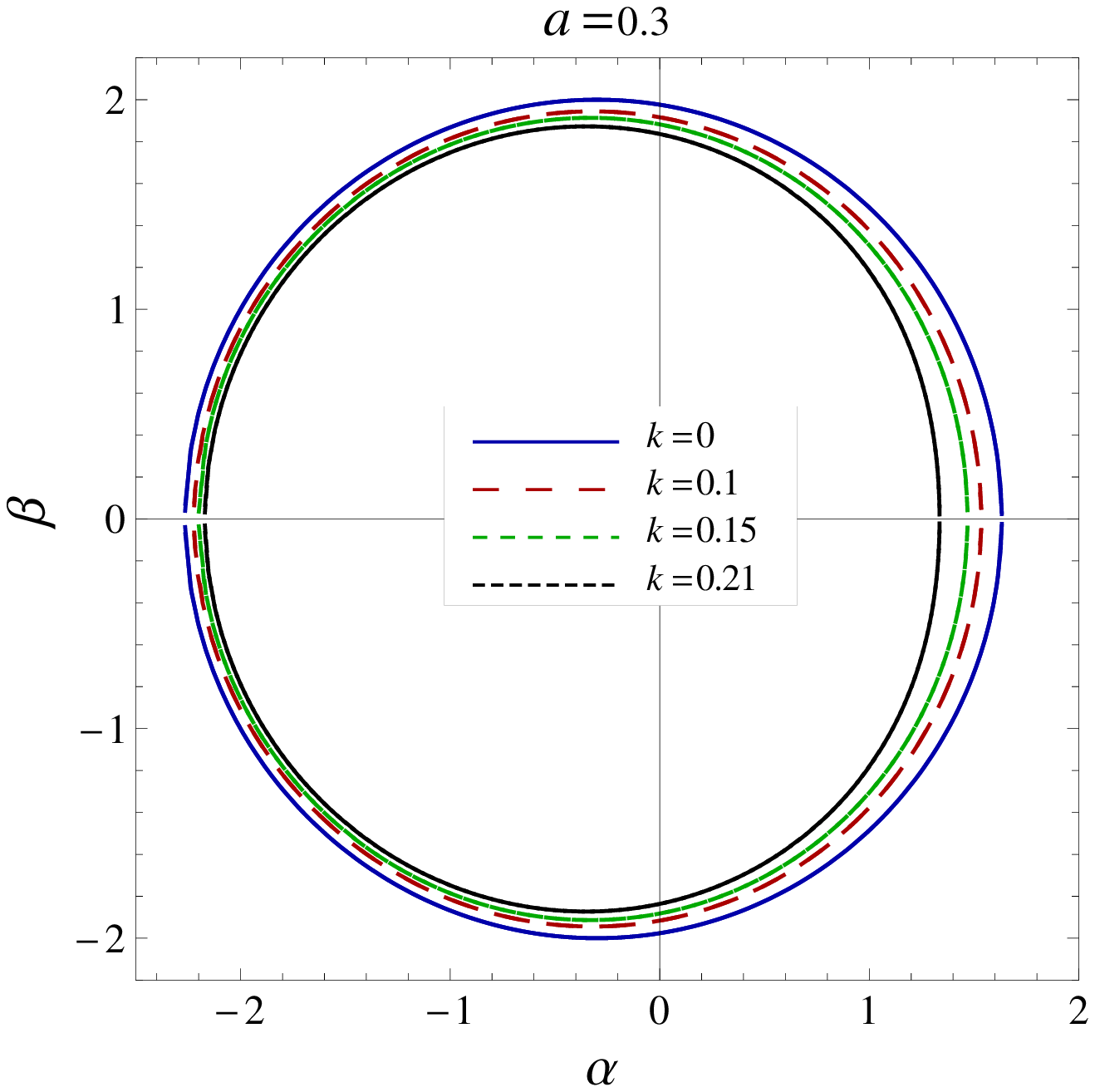}\\
\includegraphics[scale=0.31]{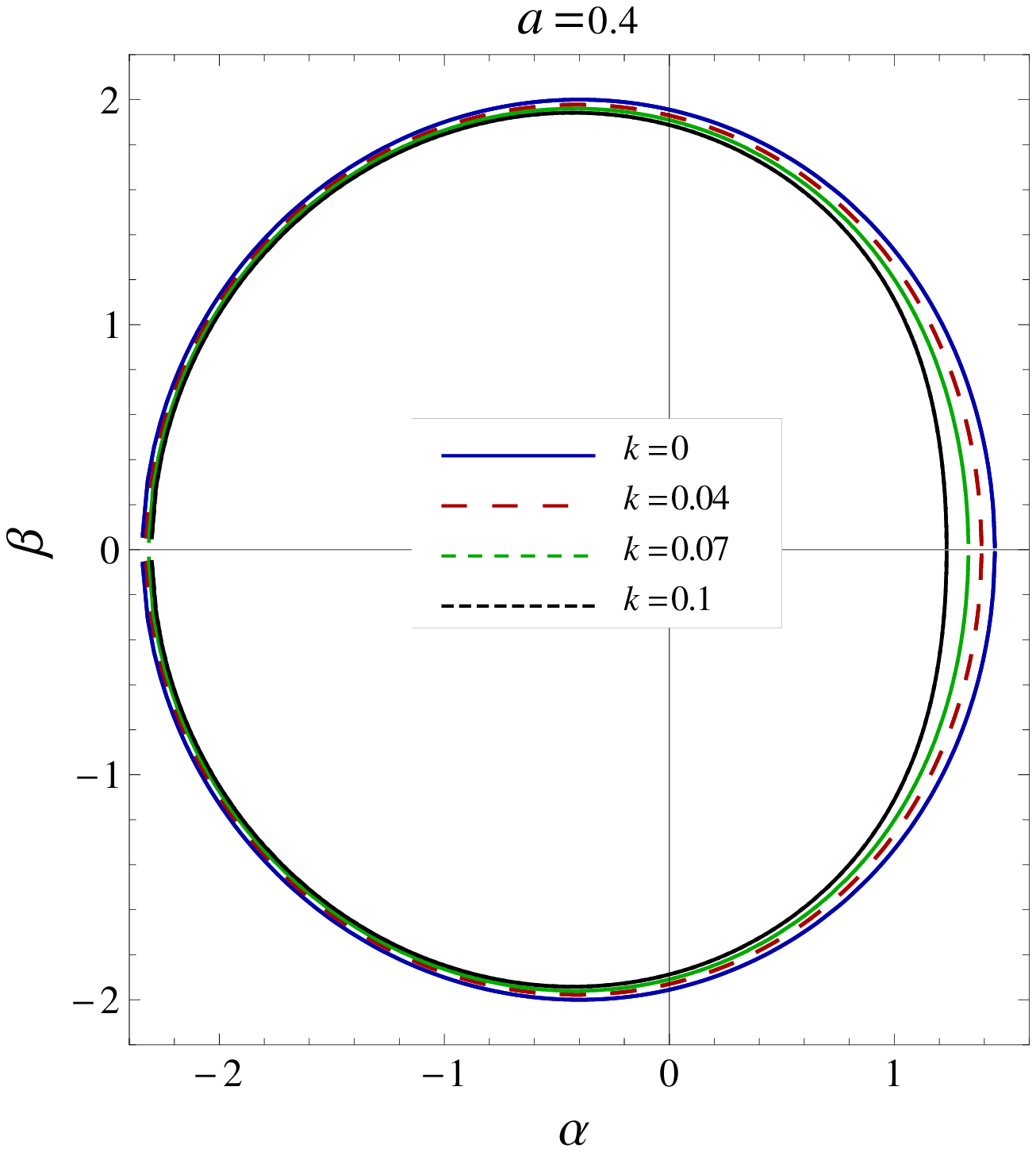}
\includegraphics[scale=0.34]{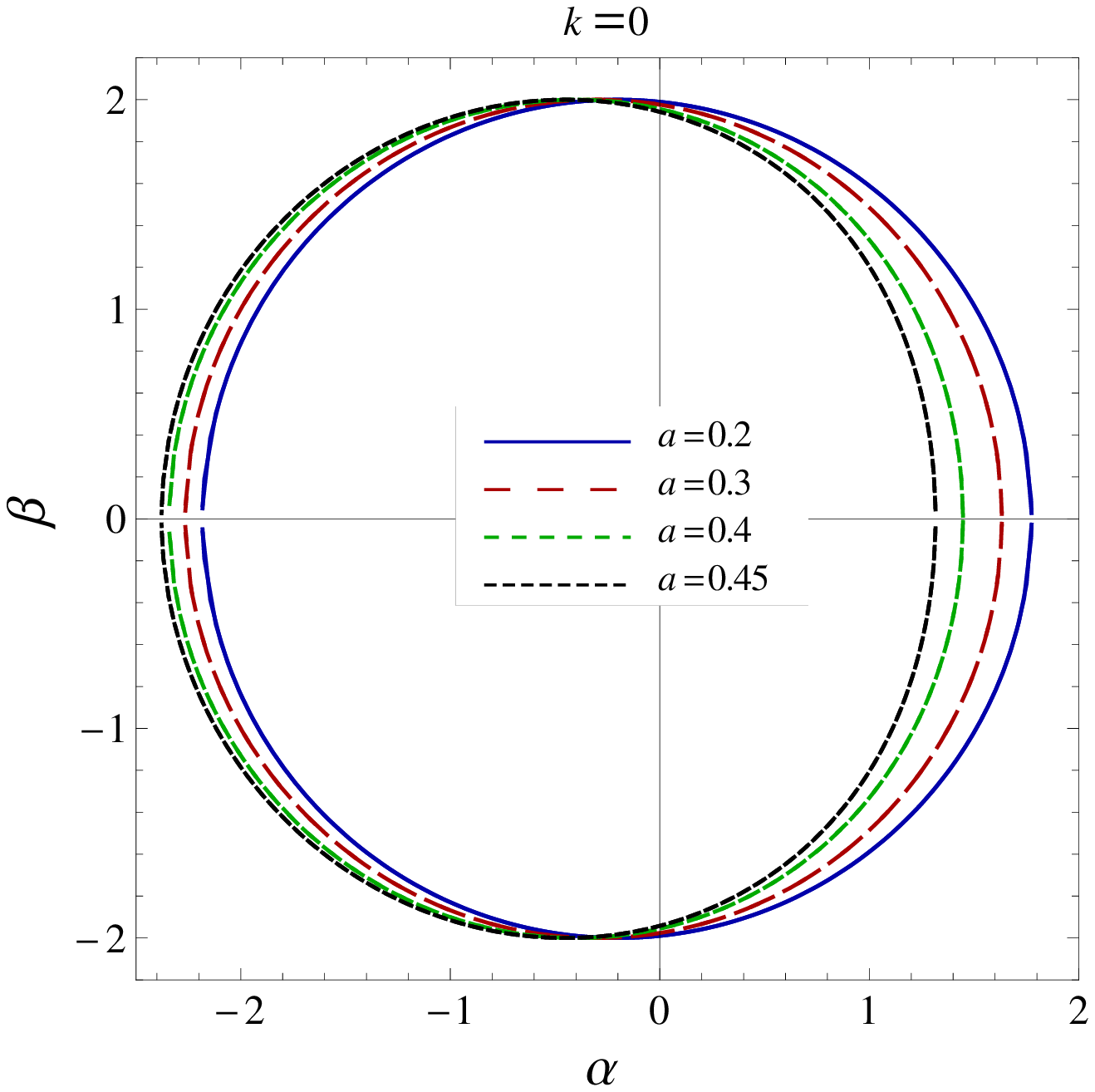}
\includegraphics[scale=0.34]{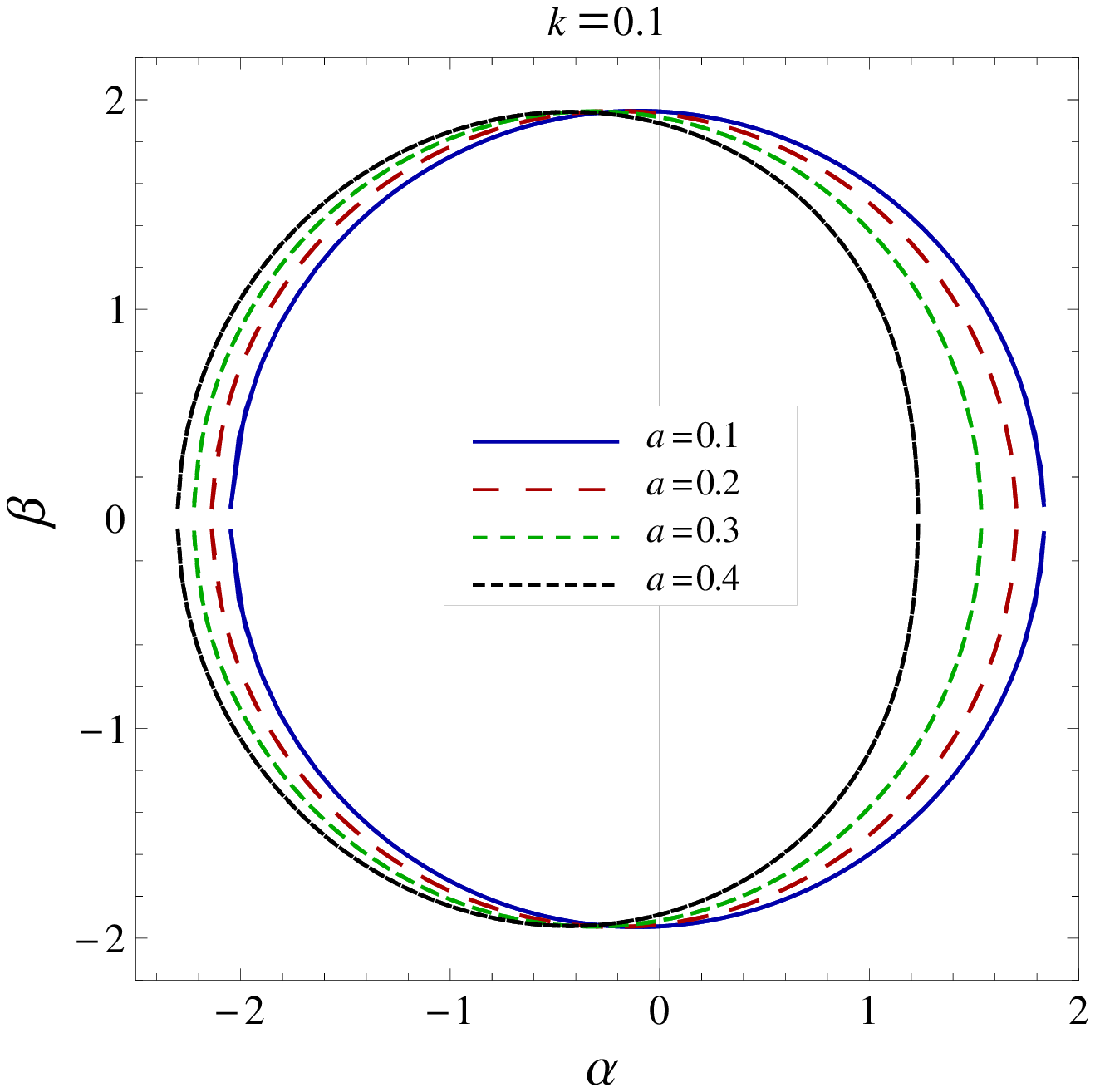}
\caption{\label{fig6} Plots showing the shapes of the black shadow for rotating regular Myers-Perry black holes with different values of $a$ and $k$.}
\end{figure*}
 In order to study the shadow of the $5D$ rotating regular Myers-Perry black hole, we use the celestial coordinates $\alpha$ and $\beta$ to determine the geometry of the shadow of the black hole, which for $5D$ black holes can be written as \cite{Amir:2017slq}
\begin{eqnarray}\label{beta}
\alpha &=& -\Big(\xi_1  \frac{1}{\sin \theta_0}+\xi_2  \frac{1}{\cos \theta_0} \Big), \nonumber\\
\beta &=& \pm \sqrt{\eta-\xi_1^2 \csc^2 \theta_0 - \xi_2^2 \sec^2 \theta_0 + a^2}.
\end{eqnarray}
\begin{figure}
	\includegraphics[scale=0.3]{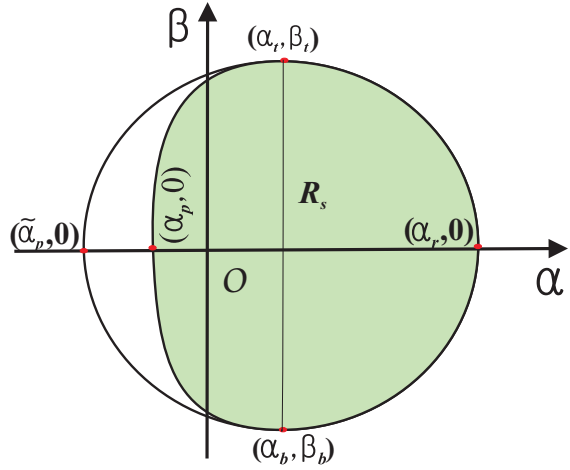}
    \caption{\label{fig7} Pictorial representation of the observables for rotating black holes
    \cite{Amir:2016cen}.}
\end{figure}
\\ As we choose an equatorial plane for observing the shadow of the black hole, so the angle of the inclination is $\theta_0 = \pi/2$. For $\theta_0 = \pi/2$, $L_{\psi}$ becomes zero as $L_{\psi}=p_{\psi}$ from Eq.~(\ref{canonical}), hence $\xi_2$ vanished. So, in this case, Eq.~(\ref{beta}) takes the form
 \begin{eqnarray}\label{alpha1}
\alpha &=& - \xi_1, \nonumber\\
\beta &=& \pm \sqrt{\eta - \xi_1^2 + a^2}.
\end{eqnarray}
The different shapes of the shadow of a $5D$ rotating regular Myers-Perry black hole can be visualized by plotting the celestial coordinates $\alpha$ vs $\beta$ for different values of the rotation parameter $a$ and deviation parameter $k$. It can also verify from Eq.~(\ref{alpha1}), that the celestial coordinates $\alpha $ and $\beta$  satisfy the relation $\alpha^2+\beta^2=\eta+ a^2$, where $\eta$ is given by Eq.~(\ref{eta}). The shadow for $5D$ non-rotating regular black hole can be obtained from
\begin{equation}\label{albe}
\alpha^2+\beta^2= \frac{4}{2-e^{-k/2}}.
\end{equation}
We plot the shadow for non-rotating case ($a=0$) in Fig.~\ref{fig5}, for different values of parameter $k$ and find that the shadow is a perfect circle and the effect of $k$ leads to decrease in size with increasing $k$.\\
Next, we analyze the behavior of black hole shadow for rotating case ($a \neq 0$) and also discuss the effect of the free parameter $k$. 
The contour of the shadow of a $5D$ rotating regular Myers-Perry black holes has been plotted in Fig.~\ref{fig6} for different values of $k$ and $a$. We realize that the shape of the contour of the shadow is largely affected due to the parameters $a$, $k$, and an extra dimension (cf. Fig.~\ref{fig6}). It is known that shadow gets more and more distorted due to increasing the rotation parameter $a$, and it is also true for the case of $5D$ rotating regular Myers-Perry black hole (cf. Fig.~\ref{fig6}). The size of the shadow decreases with increasing the value of $k$. The extra dimension also has an effect on the size of the shadow, which can be seen if we compare our present result with its four-dimensional counterpart \cite{Amir:2016cen}. 
\begin{figure*}
	\includegraphics[scale=0.52]{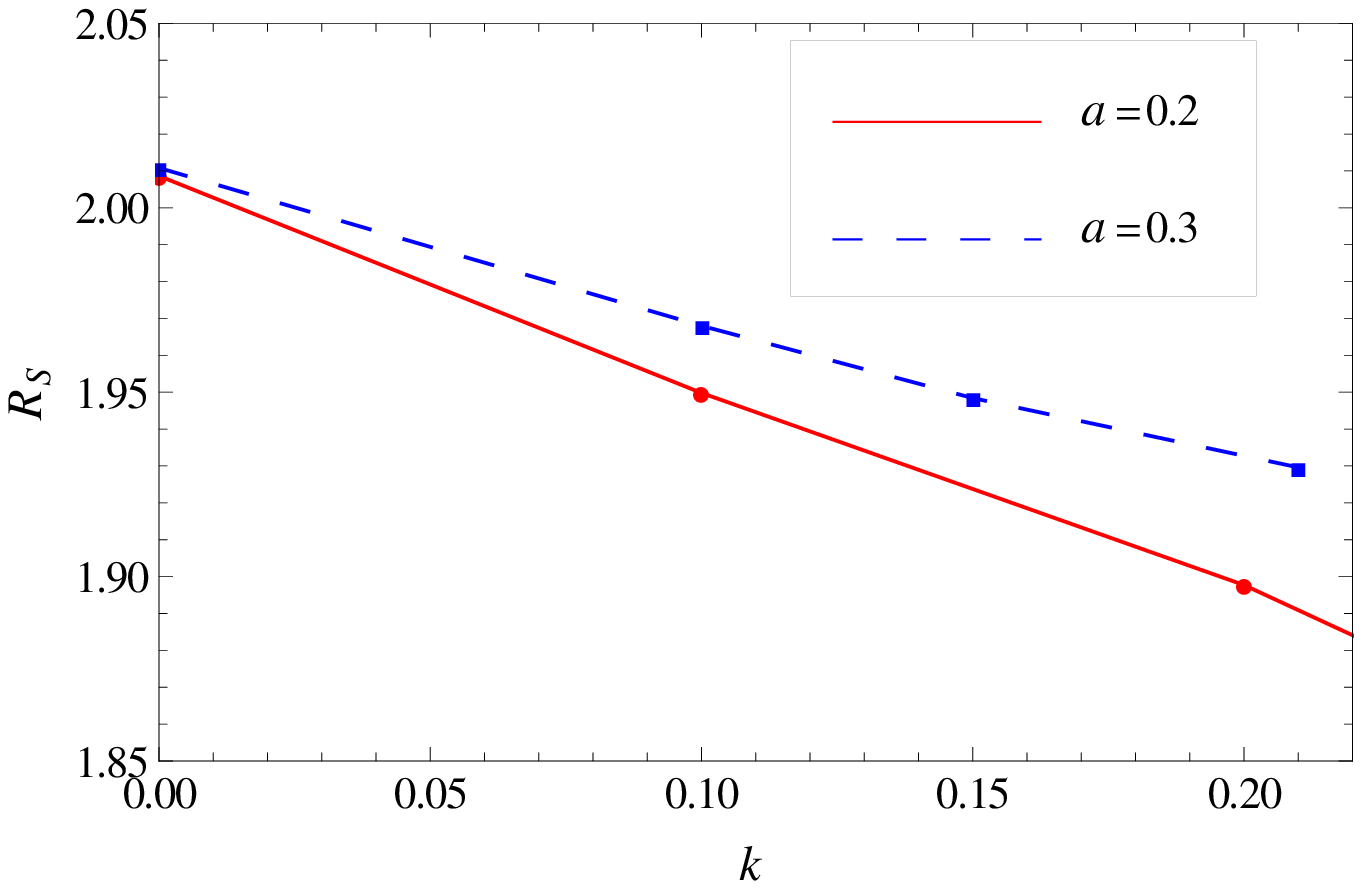}
	\includegraphics[scale=0.52]{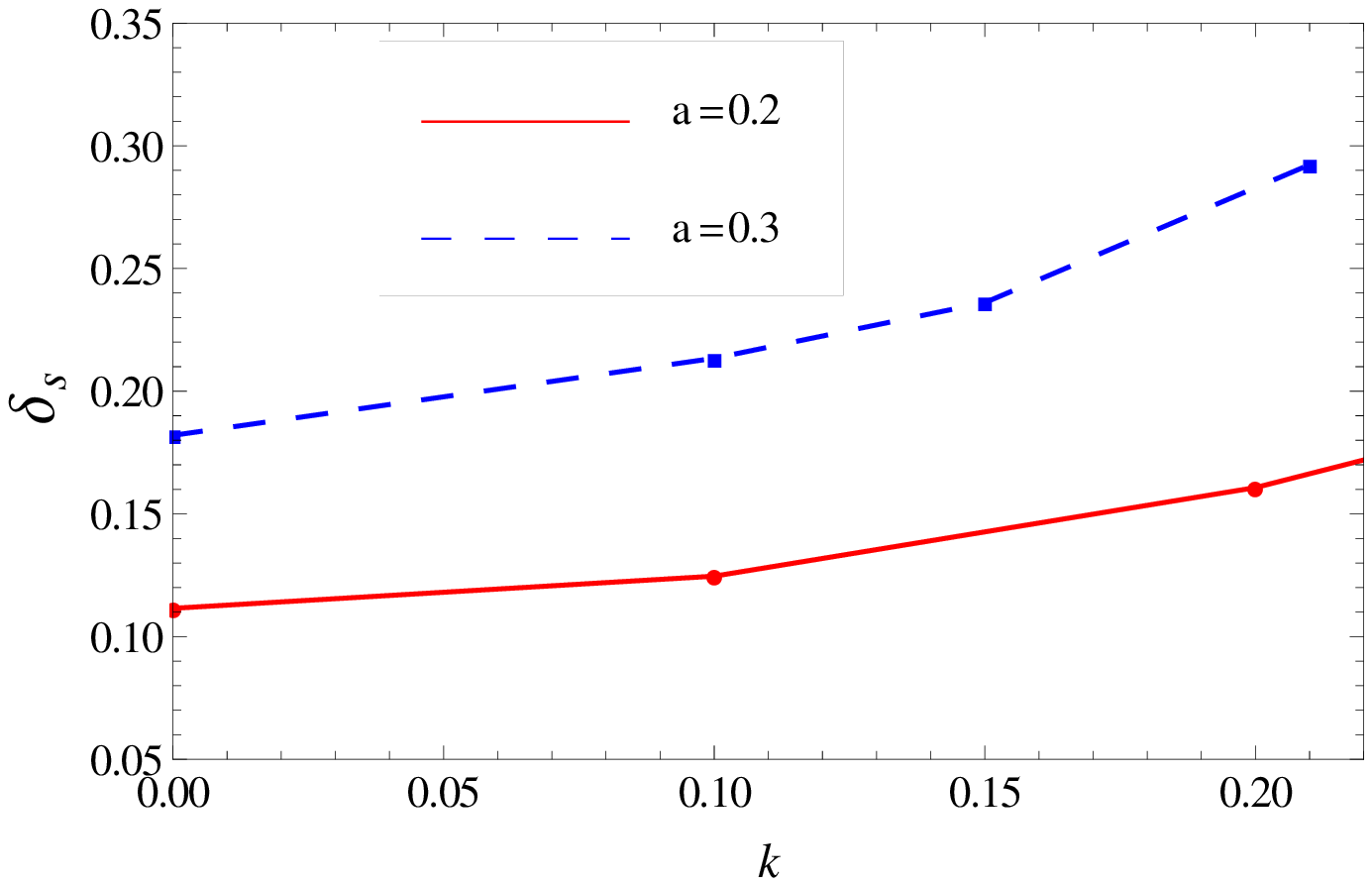}
    \caption{\label{fig8} Plots showing the variation of the radius of shadow $R_s$ and distortion parameter $\delta_s$ with parameters $k$ and $a$.}
\end{figure*}
\\To further analyze the shadow of $5D$ rotating regular Myers-Perry black hole, we adapt the two astronomical observable originally introduced by Hioki-Maeda \cite{Hioki:2009na}, which are given by
\begin{eqnarray}\label{ares}
R_s &=& \frac{(\alpha_t - \alpha_r)^2 + \beta_t^2}{2(\alpha_t - \alpha_r)}, \\
\delta_{s} &=& \frac{\tilde{\alpha_p} -\alpha_p}{R_s},
\end{eqnarray}
where $R_s$ corresponds to the size of the shadow and $\delta_{s}$ measures the deformation (cf. Fig.~\ref{fig7}). 
\\ We choose a reference circle, which passes through the three coordinates, i.e., at top position ($\alpha_t$, $\beta_t$), bottom 
position ($\alpha_b$, $\beta_b$), and rightmost position ($\alpha_r$, 0). Now we define two other points, $(\tilde{\alpha_p} ,0)$ and $(\alpha_p,0)$, which are the coordinates of the reference circle and contour of the shadow and located at diametrically opposite to ($\alpha_r$, 0) (cf. Fig.~\ref{fig7}). The radius $R_s$ and distortion $\delta_s$ for different values of parameter $k$ and rotation parameter $a$ is shown in Fig.~\ref{fig8}. We find that shadow radius $R_s$ decrease with increasing the value of $k$ and increases with rotation parameter $a$. However, the deformation $\delta_s$ increasing monotonically with $k$ as well as rotation parameter $a$. These results are consistent with our result from Figs.~\ref{fig5} and \ref{fig6}, and our earlier result \cite{Papnoi:2014aaa}.
\begin{figure*}
	\includegraphics[scale=0.55]{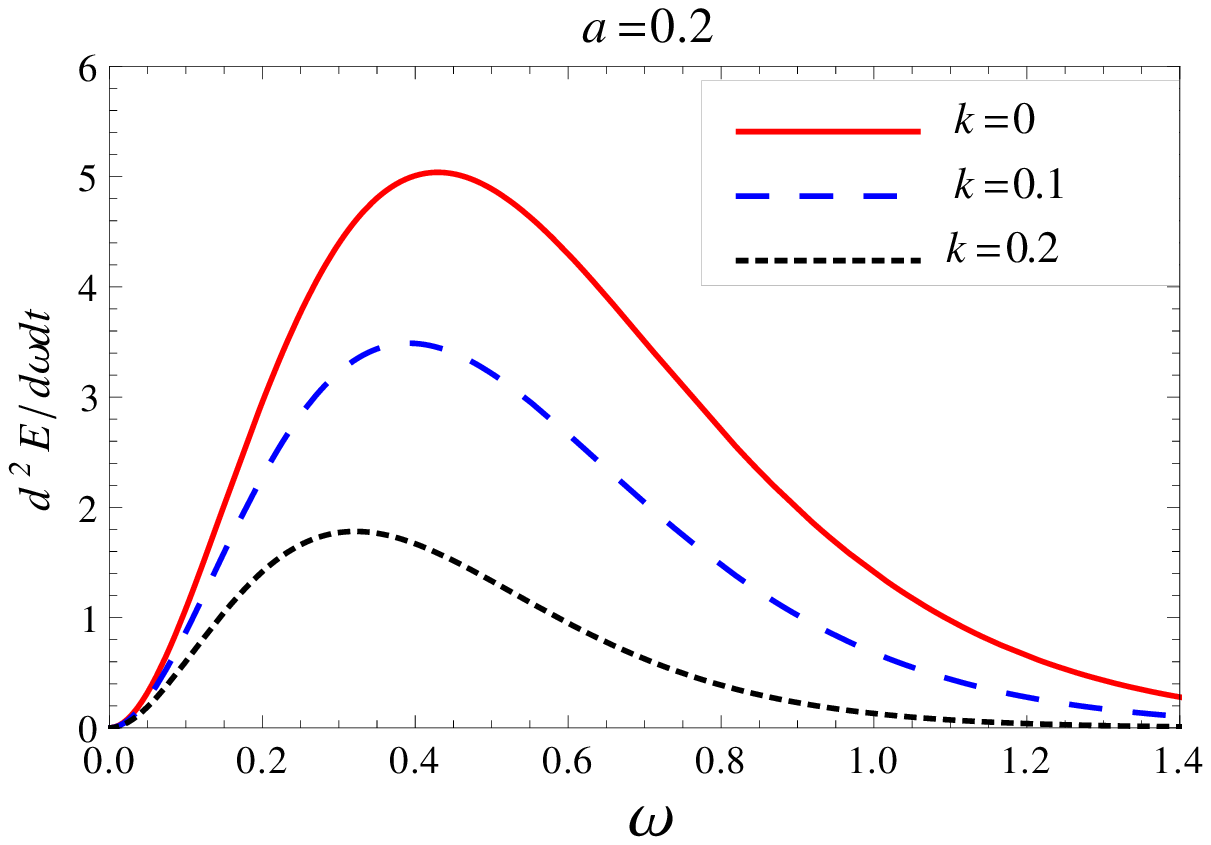}
	\includegraphics[scale=0.55]{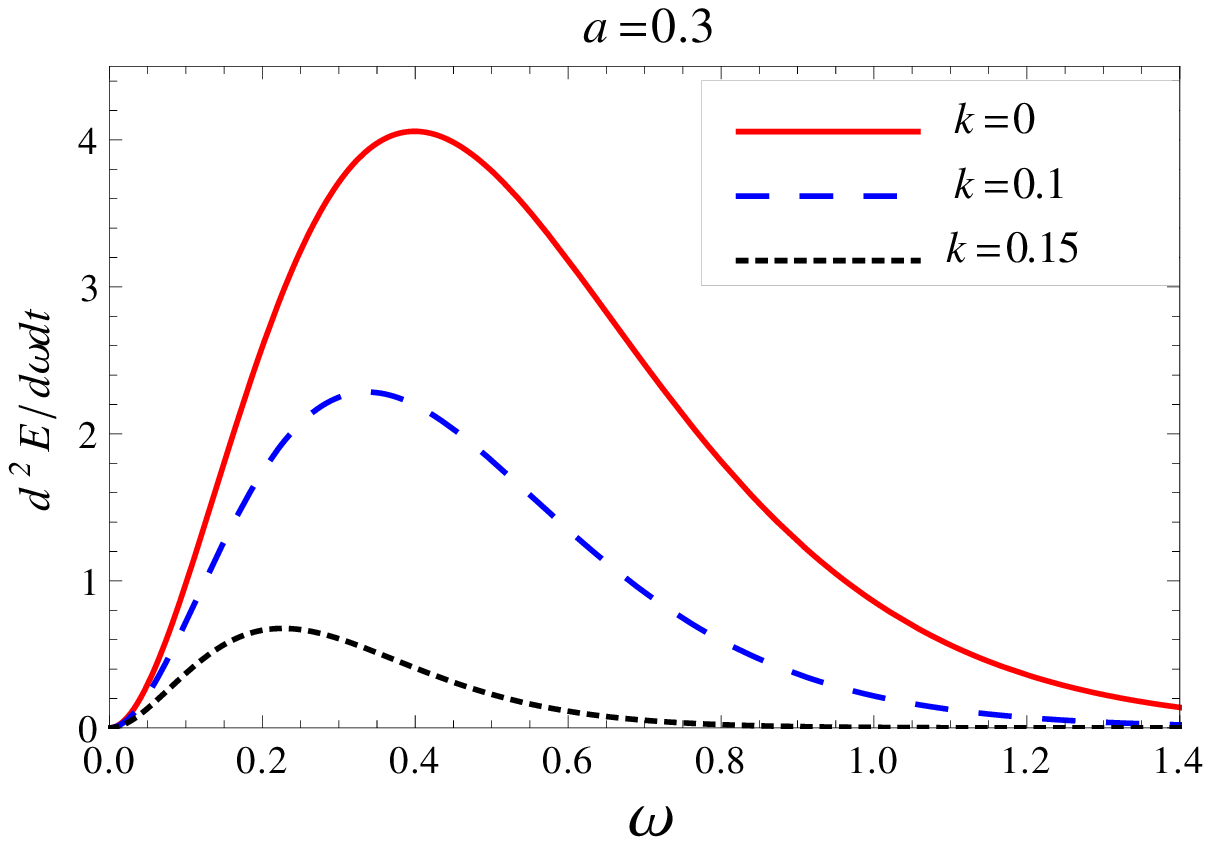}
    \caption{\label{fig9} Plots showing the variation of the energy emission rate with frequency $\omega$ for different values of parameters $k$ and $a$.}
\end{figure*}
\\ Next, we will discuss the rate of energy emission for the $5D$ rotating regular Myers-Perry black hole (\ref{metric5}). The low energy absorption cross-section for a spherically symmetric black hole has a universal feature in that it always reduces to the black hole horizon area \cite{Das:1996we,Decanini:2011xi}. However, at high energy scale, the absorption cross section oscillates around a limiting constant value which took the value of geometrical cross section $\sigma_{lim}$ of the photon sphere in which the black hole is endowed and takes the value
\begin{equation}
\sigma_{lim}=\pi R_s^3,
\end{equation}
where $R_s$ is black hole shadow radius. It is believed that it is a generic feature of a black hole at high energy surrounded by photon sphere. The expression of the energy emission rate of black hole reads \cite{Mashhoon:1973zz}
\begin{equation}\label{enem}
\frac{d^2E{\nu}}{d\nu dt} = \frac{4 \pi^3 R_s^3 }{e^{\omega/T_{+}}-1}\omega^3 = \frac{4 \pi^2 \sigma_{lim}}{e^{\omega/T_{+}}-1}\omega^3,
\end{equation}
where $\omega$ is the frequency of a photon and $T_{+}$ is the Hawking temperature on the event horizon, which can be calculated via definition
\begin{eqnarray}
T_{+}=\lim_{r \rightarrow r_+} \frac{\partial_r \sqrt{g_{tt}}}{2 \pi \sqrt{g_{rr}}},
\end{eqnarray}
Thus the temperature is
\begin{eqnarray}\label{temp}
T_{+}=\frac{r_{+}^2(r_{+}^2-a^2)-k (r_{+}^2+a^2)}{2 \pi r_{+}^3 (r_{+}^2+a^2)}.
\end{eqnarray}
If $k=0$, then Eq.~(\ref{temp}) reduces to the temperature of $5D$ Myers-Perry black hole \cite{Pourdarvish:2014vda}. We show energy emission rate in Fig.~\ref{fig9}, against the frequency $\omega$ for different values of the parameter $k$. We find that the peak of emission rate monotonically decreases and also shifts to left, i.e., to lower frequency. 

\section{Conclusion}\label{cnbh}
In this paper, we find an exact $5D$ spherical symmetric regular black hole by coupling general relativity with nonlinear electrodynamics for an appropriate lagrangian. In turn, we construct $5D$ rotating regular black hole with two angular momenta using modified Newman-Janis algorithm \cite{Erbin:2014lwa}. The $5D$ rotating regular black hole is characterized by parameters mass ($M$), two rotation parameters ($a,\;b$), and an additional deviation parameter ($k$) due to magnetic charge. It encompasses the $5D$ Myers-Perry black hole as a special case in the absence of charge. Interestingly, the parameter $k>0$ makes a significant impact on the horizons and ergoregions of the black hole. It turns out that the ergoregion area increases with the inclusion of the parameter $k$ (cf. Fig.~\ref{ergo}). Indeed, one can find a critical $k=k_c$, for a set of values other parameters, where two horizons merge and no horizons for $k>k_c$ (cf. Fig.~\ref{ehf}).\\
A black hole casts a shadow when situated in front of a bright object. We have analyzed the shadow cast by $5D$ rotating regular Myers-Perry black hole. Recent years  witnessed flurry of activities in theoretical investigation of black hole shadow because of the possibility of imaging the Sgr $\mbox{A}^{\star}$, the black hole in the center of our galaxy, by the Event Horizon telescope. The shadow of a black hole may be very useful to understand the gravity near extreme region. In the Kerr black hole case, the apparent shape of shadow gets distorted when black hole rotates very fast (for $a \geq 0.6$), whereas the size for the black hole decreases in the case of Kerr-Newman, i.e, when one introduces charge.  We have analyze how the shadow of the black hole gets affected due to parameter $k$. We have the shadows cast by the $5D$ regular Myers-Perry black hole for both rotating and non-rotating case. It is shown that null geodesics, as in the Kerr black hole, are separable. Despite of complexity, we have analyzed photon orbits around the black hole to obtain exact expressions for shadow observable $R_s$ and $\delta_s$. From the analysis of $R_s$, the shadow of the $5D$ rotating regular Myers-Perry black hole decreases with increasing value of magnetic charge parameter $k$, as in the case of Kerr-Newman black hole \cite{Vries}. The distortion of shadow is characterized by the observable $\delta_s$, which monotonically increases with increasing value of parameter $k$, i.e., the $5D$ rotating regular Myers-Perry black hole is more distorted.  Our analysis gives rich properties of the $5D$ rotating regular Myers-Perry black hole.
    
 \section{Acknowledgement}
S.G.G. would like to thanks DST INDO-SA bilateral project DST/INT/South Africa/P-06/2016.  

\end{document}